\documentclass[english,aps,prb,superscriptaddress,reprint,longbibliograhy]{revtex4-2}
\usepackage[T1]{fontenc}
\usepackage[latin9]{inputenc}
\usepackage{babel}
\usepackage{graphicx}
\usepackage{bm}
\usepackage{amsmath}
\usepackage{amssymb}
\usepackage{amsfonts}
\usepackage{dcolumn}
\usepackage{bbold}
\usepackage{tikz}
\usepackage{xfp}
\usepackage{verbatim}
\usepackage{soul}
\usepackage{ulem}
\usepackage[margin=15mm]{geometry}
\colorlet{linkequation}{blue}
\usepackage[colorlinks=true,linkcolor=blue,citecolor=red,urlcolor=blue]{hyperref}
\usetikzlibrary{shadings}

\usepackage{color}
\begin{document}

%----------------------------------------------------------------------
\begin{abstract}

The temperature dependence of mesoscopic spin-model parameters is derived in two-sublattice antiferromagnetically aligned systems based on Green's function theory. It is found that transversal spin correlations decrease the anisotropy terms while increasing the Heisenberg and Dzyaloshinsky--Moriya exchange interactions and the latter's contribution to the anisotropy. The obtained temperature dependences show quantitative agreement with the results for ferromagnets, and they also agree well with numerical atomistic simulations which treat the spin correlations without approximations. Possible applications of the results in multiscale modelling are discussed.

\end{abstract}
%----------------------------------------------------------------------

\title{Temperature dependence of spin-model parameters in antiferromagnets}

\author{Levente R\'ozsa}
\email[]{levente.rozsa@uni-konstanz.de}
\affiliation{Department of Physics, University of Konstanz, DE-78457 Konstanz, Germany} 
\affiliation{Department of Theoretical Solid State Physics, Institute of Solid State Physics and Optics, Wigner Research Centre for Physics, H-1525 Budapest, Hungary}
\affiliation{Department of Theoretical Physics, Budapest University of Technology and Economics, H-1111 Budapest, Hungary}

\author{Unai Atxitia}
\affiliation{Instituto de Ciencia de Materiales de Madrid, CSIC, Cantoblanco 28049 Madrid, Spain} 

\maketitle
\date{\today}

\section{Introduction}

The most widely studied class of antiferromagnets contains two sublattices on which the magnetic moments point oppositely to each other. Materials where the magnitude of the moments on the sublattices is different are known as ferrimagnets. Both antiferromagnets~\cite{Gomonay2014,Jungwirth2016,Baltz2018,Nemec2018} and ferrimagnets~\cite{Barker2021,Kim2022}
have attracted much attention recently as a material platform for spintronics.
Their dynamics is typically orders of magnitude faster than that of ferromagnets, including higher spin-wave frequencies~\cite{Gomonay2014}; relativistic domain wall motion \cite{Otxoa2020,Caretta2020}; enhanced magnetization switching rates induced by current~\cite{Wadley2016}, thermal excitations~\cite{Meinert2018,Rozsa2} or ultrashort laser pulses~\cite{Ostler2012,Jakobs2022a,Jakobs2022c}; and increased demagnetization speeds~\cite{Jakobs2022b}. 
Antiferromagnets may transform into a phase where the sublattice magnetizations have a weak parallel or ferromagnetic component. This transformation can be achieved by applying an external field~\cite{Tsujikawa1959}, increasing the temperature~\cite{Morin1950}, or even in the ground state in the presence of the Dzyaloshinsky--Moriya interaction~\cite{Dzyaloshinsky1958,Moriya1960}. 
Due to the different temperature dependence of the sublattice magnetizations and angular momenta, they can become compensated in certain ferrimagnets, 
influencing the velocity and the movement direction of domain walls and skyrmions driven by spin-polarized currents or thermal gradients~\cite{Caretta2018,Hirata2019,Donges2020}.
At the angular momentum compensation point, ferrimagnets combine the advantages of both ferromagnets and antiferromagnets: easy control and detection of their net magnetization by an external field, antiferromagnetic-like ultrafast dynamics and the potential for high-density devices.

These phenomena can often be successfully modelled by theoretical approaches agnostic to the underlying atomic structure, such as finite-temperature macrospin models like the Landau--Lifshitz--Bloch equation~\cite{Atxitia_2017} or continuum theories. In these models, the on-site anisotropy contributions and the pair-wise interactions, such as exchange or Dzyaloshinsky--Moriya terms, are intrinsically temperature dependent due to averaging over the fluctuations and correlations of atomic spins in a finite volume. 
Computer simulations using atomistic spin models~\cite{Nowak2007BOOK} can naturally describe the equilibrium thermodynamics and non-equilibrium dynamics of antiferromagnets and ferrimagnets, but they are considerably more resource intensive.
The price paid for the reduced computational cost of the mesoscopic methods is that the temperature-dependent effective parameters of these models are difficult to determine. 
While first-principles methods have proven successful in calculating atomistic~\cite{LIECHTENSTEIN1987,Udvardi2003,Xiang2011} or zero-temperature continuum~\cite{Heide2008,Freimuth_2014} spin-model parameters, their application to finite-temperature mesoscopic models remains limited. The temperature dependence of the magnetocrystalline anisotropy energy has been calculated based on a disordered local moment scheme~\cite{Staunton2004,Deak2014,Patrick2018}. However, this method treats spin fluctuations on a mean-field level, since modelling correlations in density functional theory is inherently challenging. Methods for calculating exchange interactions at finite temperature have been proposed in Refs.~\cite{Bottcher2012,Szilva2013}, but the degree of spin disorder was determined from atomistic simulations in these cases. %these values cannot directly be transformed to finite-temperature mesoscopic models.
Therefore, obtaining the effective parameter values for the models require fitting to experimental data obtained in a wide temperature range which are not always available, or to the results of atomistic spin-model simulations what counteracts the reduced computational cost of the mesoscopic models.

This difficulty can be circumvented by applying analytical methods which treat correlations accurately, and can approximate these temperature-dependent parameters at low computational costs. Such methods are often based on Green's functions, where the difficulty arises in choosing the decoupling scheme, i.e., the procedure for truncating the infinite series of correlation functions of increasing order. Non-linear spin-wave theory is based on conventional diagrammatic perturbation methods developed for bosonic and fermionic systems, and is widely used for describing quantum fluctuations~\cite{Oguchi1960,Zhitomirsky2013} at low temperatures. These methods are often inaccurate for spin models at elevated temperatures; for example, they predict a finite jump in the magnetization at the critical temperature in simple ferromagnets~\cite{Bloch1962,Irkhin1999}. For accurately modelling the temperature-induced phase transitions, semi-empirical decoupling schemes for spin Green's functions have been developed by Tyablikov~\cite{Tyablikov1959}, known as the random-phase approximation, and by Callen~\cite{Callen}, which have proven to be especially accurate for low and high spin values, respectively. The method has been generalized to two-sublattice antiferromagnets by Anderson and Callen~\cite{Anderson}. The applications of these methods to two-dimensional ferromagnetic and antiferromagnetic quantum spin systems is summarized in Ref.~\cite{FROBRICH2006}, and antiferromagnets with Dzyaloshinsky--Moriya interactions have been treated within the random-phase approximation in Refs.~\cite{Tabunshchyk2005,Rutonjski2021}. These works primarily focused on the calculation of the magnetizations, the correlation functions and the magnon frequencies at finite temperatures, which can be used to describe phase transitions, but not on the connection between the atomistic and mesoscopic models. The latter topic has been investigated for ferromagnets in the classical limit of infinite spin in Refs.~\cite{Bastardis,Rozsa,Evans}, but the corresponding investigations for antiferromagnetically aligned systems seem to be lacking.

Here, we derive the temperature dependence of the effective interaction parameters in mesoscopic models of two-sublattice antiferromagnets and ferrimagnets. We extend the Green's function theory in Ref.~\cite{Anderson} by including all terms preserving rotational symmetry around the axis of the magnetizations, namely Heisenberg and Dzyaloshinsky--Moriya exchange interactions as well as single-ion and two-ion anisotropy terms, and discuss both the classical and quantum cases. A comparison with atomistic Monte Carlo simulations demonstrates the accuracy of the method in treating spin correlations at a fraction of the computational cost of the numerical simulations.

The paper is organized as follows. In Sec.~\ref{sec2a}, we present the self-consistency equations of Green's function theory, which we apply to derive the correspondence between the atomistic and mesoscopic models in Sec.~\ref{sec2b}. We discuss the scaling exponents of the effective parameters in Sec.~\ref{sec2c}. We apply the method to a square lattice in Sec.~\ref{sec3} and compare the predictions with atomistic simulations.

\section{Theory\label{sec2}}

\subsection{Green's function theory\label{sec2a}}

We consider a two-sublattice magnet described by the atomistic spin Hamiltonian
\begin{align}
\mathcal{H}=&-\frac{1}{2}\sum_{i,j,r,s}\left(J_{ij}^{rs}\boldsymbol{S}_{ir}\boldsymbol{S}_{js}+\Delta J_{ij}^{rs}S_{ir}^{z}S_{js}^{z}+\boldsymbol{D}_{ij}^{rs}\left(\boldsymbol{S}_{ir}\times\boldsymbol{S}_{js}\right)\right)\nonumber
\\
&-\sum_{i,r}K^{r}\left(S_{ir}^{z}\right)^{2}-\sum_{i,r}\mu_{r}B^{z}S_{ir}^{z}.\label{eqn1}
\end{align}
Here $r,s\in\left\{A,B\right\}$ denote the two sublattices, $J_{ij}^{rs}$ is the Heisenberg exchange interaction between atoms at sites $i$ and $j$, $\boldsymbol{D}_{ij}^{rs}$ is the Dzyaloshinsky--Moriya vector, $\Delta J_{ij}^{rs}$ is the two-ion anisotropy, $K^{r}$ is the single-ion magnetocrystalline anisotropy, $\mu_{r}$ is the magnetic moment, and $B^{z}$ is the external magnetic field. 
We note that this model describes an antiferromagnet when $\mu_A=\mu_B$ and a ferrimagnet when $\mu_A \neq \mu_B$.
$\boldsymbol{S}_{ir}$ stands for the spin vectors; for most considerations they will be treated as classical unit vectors $|\boldsymbol{S}_{ir}|=1$, since atomistic spin-model simulations used for comparison are easier to perform in the classical limit. However, at certain points the quantum-mechanical case with spin operators will be discussed.
The number of unit cells in the lattice will be denoted by $N_{\textrm{c}}$, corresponding to the number of atoms in both the $A$ and the $B$ sublattices. Note that fully analytical results in the limit $N_{\textrm{c}} \rightarrow \infty$ are only available when the types of interactions are more restricted; therefore, in most cases we will rely on semi-analytical techniques where lattice sums over a finite number of lattice sites must be performed.
It will be assumed that in the classical ground state, spins on sublattice $A$ are oriented along the $+z$ direction, while spins on sublattice $B$ point along the $-z$ direction. To stabilize the antiferromagnetic alignment, it will be assumed that the antiferromagnetic intersublattice coupling $J_{ij}^{AB}<0$ is dominant compared to the intrasublattice coupling and the Dzyaloshinsky--Moriya interaction, while the anisotropy prefers spin alignment along the $z$ axis.
%Only the $z$ component of the Dzyaloshinsky--Moriya vectors will be treated. This way, Eq.~\eqref{eqn1} contains all possible single-spin and two-spin terms that are rotationally invariant around the $z$ direction. The rotational symmetry simplifies the following calculations.

The equation of motion generated by the Hamiltonian $\mathcal{H}$ reads
\begin{eqnarray}
\partial_{t}\boldsymbol{S}_{ir}=-\gamma \boldsymbol{S}_{ir}\times\boldsymbol{B}_{ir}^{\textrm{eff}}\label{eqn1a}
\end{eqnarray}
in the classical limit, where $\gamma$ is the gyromagnetic ratio. Equation~\eqref{eqn1a} describes the precession of the spins around the effective magnetic field $\boldsymbol{B}_{ir}^{\textrm{eff}}=-\mu_{r}^{-1}\partial\mathcal{H}/\partial\boldsymbol{S}_{ir}$. We introduce a local coordinate system where all spins are oriented along the $+z$ direction in the classical ground state, with $\tilde{\boldsymbol{S}}_{iA}=\boldsymbol{S}_{iA},\tilde{S}_{iB}^{z}=-S_{iB}^{z}$ and $\tilde{S}_{iB}^{\pm}=-S_{iB}^{\mp}$, where $S_{iB}^{\pm}=S_{iB}^{x}\pm\textrm{i}S_{iB}^{y}$ denotes the ladder operators. In linear spin-wave theory, the dynamical equation is linearized around the classical ground state in the quantities $S_{ir}^{\pm}$. We introduce the shorter notations $\tilde{S}_{ir}^{(1)}\in\left\{\tilde{S}_{iA}^{+},\tilde{S}_{iB}^{-}\right\}$ and $\tilde{S}_{ir}^{(2)}\in\left\{\tilde{S}_{iA}^{-},\tilde{S}_{iB}^{+}\right\}$ since this linearized equation couples the $+$ and $-$ indices between the $A$ and $B$ sublattices. After performing spatial and temporal Fourier transformation $\partial_{t}\rightarrow -\textrm{i}\omega$, we will use the notations
\begin{eqnarray}
\tilde{\boldsymbol{S}}_{\boldsymbol{q}r}&=&\frac{1}{\sqrt{N_{\textrm{c}}}}\sum_{\boldsymbol{R}_{i}}\textrm{e}^{-\textrm{i}\boldsymbol{q}\left(\boldsymbol{R}_{i}-\boldsymbol{R}_{j}\right)}\tilde{\boldsymbol{S}}_{ir},\label{eqn1b}
\\
\mathfrak{J}_{\boldsymbol{q}}^{rs}&=&\sum_{\boldsymbol{R}_{i}-\boldsymbol{R}_{j}}\textrm{e}^{-\textrm{i}\boldsymbol{q}\left(\boldsymbol{R}_{i}-\boldsymbol{R}_{j}\right)}\left(J_{ij}^{rs}+\Delta J_{ij}^{rs}+2K^{r}\delta^{rs}\right),\label{eqn1c}
\\
\mathfrak{J}_{\boldsymbol{q}}^{'rs}&=&\sum_{\boldsymbol{R}_{i}-\boldsymbol{R}_{j}}\textrm{e}^{-\textrm{i}\boldsymbol{q}\left(\boldsymbol{R}_{i}-\boldsymbol{R}_{j}\right)}\left(J_{ij}^{rs}+\textrm{i}D_{ij}^{z,rs}\right).\label{eqn1d}
\end{eqnarray}
Here, $\boldsymbol{R}_i-\boldsymbol{R}_j$ denotes the vector connecting the lattice sites $i$ and $j$, where $\boldsymbol{R}_i = (x_i, y_i, z_i)$ stands for the position of the spin $i$ in the lattice. Note that only the $z$ component of the Dzyaloshinsky--Moriya vectors appears in these expressions. In these variables, the linearized equation of motion reads
\begin{eqnarray}
\omega \underline{S}_{\boldsymbol{q}}^{(2)}=\gamma\underline{\underline{\mu}}^{-1}\underline{\underline{\tilde{H}}}_{\textrm{SW},\boldsymbol{q}}\underline{S}_{\boldsymbol{q}}^{(2)},\label{eqn1e}
\end{eqnarray}
where single and double underlines denote vectors and matrices in sublattice indices $r,s$. Here $\underline{\underline{\mu}}=\textrm{diag}\left(\underline{\mu}\right)$ and the spin-wave Hamiltonian is
\begin{eqnarray}
\underline{\underline{\tilde{H}}}_{\textrm{SW},\boldsymbol{q}}=\textrm{diag}\left(\underline{\underline{\mathfrak{J}}}_{\boldsymbol{0}}\underline{\sigma}+\underline{\mu}B^{z}\right)-\underline{\underline{\mathfrak{J}}}_{\boldsymbol{q}}^{'}\underline{\underline{\sigma}}^{z},\label{eqn1f}
\end{eqnarray}
with $\underline{\sigma}=\left[1,-1\right]^{T}$ and $\underline{\underline{\sigma}}^{z}=\textrm{diag}\left(\underline{\sigma}\right)$ a Pauli matrix, which appear due to the antiparallel alignment of the sublattices.

Note that the $(1)$ and $(2)$ components of the spins are decoupled in the linearized equation of motion, because the Hamiltonian in Eq.~\eqref{eqn1} contains all possible single-spin and two-spin terms that are rotationally invariant around the $z$ direction.  This simplification may be justified in systems with at least a threefold rotational symmetry around the N\'{e}el vector, such as the trigonal antiferromagnet $\alpha$-Fe$_{2}$O$_{3}$~\cite{Morin1950}, hexagonal ferrimagnets like GdCo$_{5}$~\cite{Nassau1960} and cubic ferrimagnets including Mn$_{2}$Ru$_{x}$Ga~\cite{Kurt2014}. Even if the rotational symmetry is lower, such as in atomically thin antiferromagnetic Mn layers on Nb$(110)$~\cite{LoConte2022}, this model should provide a useful approximation if the anisotropy terms are considerably weaker than the exchange interactions.

The eigenvalues of Eq.~\eqref{eqn1e} correspond to the magnon frequencies. The thermal occupation of the magnon modes enables the calculation of the sublattice magnetizations and the two-spin correlation functions at low temperatures~\cite{Anderson1952}. At elevated temperatures, a higher number of magnons becomes excited, leading to a temperature-dependent renormalization of the frequencies. A self-consistent procedure for treating this renormalization based on Green's functions was introduced in Refs.~\cite{Callen,Anderson}, which we apply to the present system here. Details of the derivation are given in Appendix~\ref{appendixA}. In this method, the single-particle excitation spectrum is given by the eigenvalues of the matrix $\gamma\underline{\underline{\mu}}^{-1}\underline{\underline{\tilde{\Gamma}}}_{\boldsymbol{q}}$, where the matrix $\underline{\underline{\tilde{\Gamma}}}_{\boldsymbol{q}}$ replacing the spin-wave Hamiltonian $\underline{\underline{\tilde{H}}}_{\textrm{SW},\boldsymbol{q}}$ in Eq.~\eqref{eqn1f} reads 
\begin{align}
&\underline{\underline{\tilde{\Gamma}}}_{\boldsymbol{q}}=\textrm{diag}\Bigg(\left[\underline{\underline{\mathfrak{J}}}_{\boldsymbol{0}}\left<\underline{\underline{\tilde{S}}}^{z}\right>+2\alpha_{0}\sum_{\boldsymbol{q}'}\left(\left<\underline{\underline{\tilde{S}}}^{z}\right>\underline{\underline{\mathfrak{J}}}_{\boldsymbol{q}}^{'}\left<\underline{\underline{\tilde{S}}}^{z}\right>\right)\circ\underline{\underline{\Phi}}_{\boldsymbol{q}'}\right]\underline{\sigma}\nonumber
\\
&+\underline{\mu}B^{z}\Bigg)-\left[\left<\underline{\underline{\tilde{S}}}^{z}\right>\underline{\underline{\mathfrak{J}}}_{\boldsymbol{q}}^{'}+2\alpha_{0}\sum_{\boldsymbol{q}'}\left(\left<\underline{\underline{\tilde{S}}}^{z}\right>\underline{\underline{\mathfrak{J}}}_{\boldsymbol{q}-\boldsymbol{q}'}\left<\underline{\underline{\tilde{S}}}^{z}\right>\right)\circ\underline{\underline{\Phi}}_{\boldsymbol{q}'}^{T}\right]\underline{\underline{\sigma}}^{z},\label{eqn1g}
\end{align}
where $\left<\underline{\underline{\tilde{S}}}^{z}\right>=\textrm{diag}\left(\left<\underline{\tilde{S}}^{z}\right>\right)$ and $\circ$ denotes element-wise multiplication. The coefficient $\alpha_{0}$ is a phenomenological constant required for the decoupling of the Green's functions (see Appendix~\ref{appendixA}); here it will be set to $\alpha_{0}=1/2$ in the classical limit of the decoupling scheme proposed by Callen and Anderson~\cite{Callen,Anderson}. The $\underline{\underline{\Phi}}_{\boldsymbol{q}}$ matrix is related to the transversal spin correlation functions via
\begin{eqnarray}
\left<\underline{\tilde{S}}^{(1)}\left(\underline{\tilde{S}}^{(2)}\right)^{T}\right>=2\left<\underline{\underline{\tilde{S}}}^{z}\right>\underline{\underline{\Phi}}_{\boldsymbol{q}}.\label{eqn1h}
\end{eqnarray}
The magnon frequencies are expressed as
\begin{align}
\omega^{\pm}_{\boldsymbol{q}}=&\frac{1}{2}\left(\frac{\gamma}{\mu_{A}}\tilde{\Gamma}^{AA}_{\boldsymbol{q}}+\frac{\gamma}{\mu_{B}}\tilde{\Gamma}^{BB}_{\boldsymbol{q}}\right)\pm \frac{1}{2}\left(\frac{\gamma}{\mu_{A}}\tilde{\Gamma}^{AA}_{\boldsymbol{q}}-\frac{\gamma}{\mu_{B}}\tilde{\Gamma}^{BB}_{\boldsymbol{q}}\right)\nu_{\boldsymbol{q}},\label{eqn13}
\\
\nu_{\boldsymbol{q}}=&\sqrt{1+\frac{4\frac{\gamma}{\mu_{A}}\tilde{\Gamma}^{AB}_{\boldsymbol{q}}\frac{\gamma}{\mu_{B}}\tilde{\Gamma}^{BA}_{\boldsymbol{q}}}{\left(\frac{\gamma}{\mu_{A}}\tilde{\Gamma}^{AA}_{\boldsymbol{q}}-\frac{\gamma}{\mu_{B}}\tilde{\Gamma}^{BB}_{\boldsymbol{q}}\right)^{2}}}.\label{eqn12}
\end{align}
Note that $\omega^{+}_{\boldsymbol{q}}\ge 0$ and $\omega^{-}_{\boldsymbol{q}}\le 0$ if the antiferromagnetic alignment of the sublattices is stable. For example, in the antiferromagnetic limit with identical sublattices and no external field, one obtains $\gamma\mu_{A}^{-1}\tilde{\Gamma}^{AA}_{\boldsymbol{q}}=-\gamma\mu_{B}^{-1}\tilde{\Gamma}^{BB}_{\boldsymbol{q}}$, yielding $\omega^{-}_{\boldsymbol{q}}=-\omega^{+}_{\boldsymbol{q}}$. The opposite signs of the frequencies describe that the modes have opposite circular polarizations, which has also been demonstrated experimentally in a ferromagnet recently~\cite{Nambu2020}.

Equation~\eqref{eqn1g} is made self-consistent by determining the correlation functions from the spectral theorem~\cite{FROBRICH2006}
\begin{eqnarray}
\underline{\underline{\Phi}}_{\boldsymbol{q}}=\lim_{\varepsilon\rightarrow 0+}-\textrm{Im}\int_{-\infty}^{\infty}\frac{1}{\pi N_{\textrm{c}}}\underline{\underline{\sigma}}^{z}\left(\omega+\textrm{i}\varepsilon-\gamma\underline{\underline{\mu}}^{-1}\underline{\underline{\tilde{\Gamma}}}_{\boldsymbol{q}}\right)^{-1}\gamma\underline{\underline{\mu}}^{-1}n\left(\omega\right)\textrm{d}\omega.\label{eqn1i}
\end{eqnarray}
 This integral is evaluated in Appendix~\ref{appendixA}. The function $n\left(\omega\right)=k_{\textrm{B}}T/\omega$ corresponds to the occupation number of the magnon mode with frequency $\omega$ in the classical limit in units of action. Substituting this function simplifies Eq.~\eqref{eqn1i} to
\begin{eqnarray}
\underline{\underline{\Phi}}_{\boldsymbol{q}}=\frac{1}{N_{\textrm{c}}}k_{\textrm{B}}T\underline{\underline{\sigma}}^{z}\left(\underline{\underline{\tilde{\Gamma}}}_{\boldsymbol{q}}^{-1}\right)^{T}.\label{eqn1j}
\end{eqnarray}
The sublattice magnetizations are given by the Langevin function
\begin{eqnarray}
\left<\tilde{S}_{r}^{z}\right>=\textrm{coth}\:\Phi_{r}^{-1}-\Phi_{r},\label{eqn18}
\end{eqnarray}
where $\Phi_{r}=\sum_{\boldsymbol{q}}\Phi^{rr}_{\boldsymbol{q}}$ may be interpreted as the total spin carried by the magnons on sublattice $r$. At zero temperature, $\Phi_{r}=0$ holds as can be seen from Eq.~\eqref{eqn1j}, and the sublattice magnetizations are saturated $\left<\tilde{S}_{r}^{z}\right>=1$.

It is worth noting that $\left<\tilde{S}_{r}^{z}\right>=1$ does not hold in the quantum case even for $T=0$, as is already known from linear spin-wave theory~\cite{Anderson1952}. Using spin operators in the quantum case, Eqs.~\eqref{eqn1g} and \eqref{eqn1i} remain valid, but the function $n^{\textrm{quantum}}\left(\omega\right)=\hbar \left(\textrm{e}^{\hbar\omega/\left(k_{\textrm{B}}T\right)}-1\right)^{-1}$ now gives the Bose--Einstein occupation number for $\omega>0$. The magnetic moments $\mu_{r}$ have to be replaced by $g\mu_{\textrm{B}}$, where $g$ is the spin gyromagnetic factor of the electron and $\mu_{\textrm{B}}$ is the Bohr magneton, leading to $\gamma/\mu_{r}=\hbar^{-1}$, since the magnitude of the moments is described by the spin quantum number $S$ in this case. Due to this different normalization, the decoupling coefficient reads $\alpha_{0}=1/\left(2S^{2}\right)$ for quantum spins. On the left-hand side of Eq.~\eqref{eqn1h}, the product of the spin components $\left<\tilde{S}_{-\boldsymbol{q}r}^{(1)}\tilde{S}_{\boldsymbol{q}s}^{(2)}\right>$ has to be replaced by the anticommutator $\frac{1}{2}\left<\left[\tilde{S}_{-\boldsymbol{q}r}^{(1)},\tilde{S}_{\boldsymbol{q}s}^{(2)}\right]_{+}\right>$. The expectation values of the sublattice magnetizations can be calculated from the Brillouin function as
\begin{eqnarray}
\left<\tilde{S}_{r}^{z}\right>^{\textrm{quantum}}=SB_{S}\left(SX_{r}\right),\label{eqn19}
\end{eqnarray}
with $X_{r}=2\:\textrm{arcoth}\left(2\Phi_{r}\right)$ using the definition of $\Phi_{r}$ given above. Although the notations are different, it can be shown that the system of equations presented here is equivalent to Ref.~\cite{Anderson} when the Dzyaloshinsky--Moriya and two-ion anisotropy terms are set to zero. For $T=0$, one obtains $\Phi_{r}>0$ and $\left<\tilde{S}_{r}^{z}\right><S$, indicating that the classical ground state is not the correct quantum ground state. 
Note that although the Brillouin function and in the classical limit the Langevin function also define the magnetization in mean-field theory, the argument of the functions differs from the mean-field model in Green's function theory; see Ref.~\cite{Callen3} for a detailed discussion.

The main result of the Green's function formalism is the calculation of the frequencies of the two magnon modes $\omega^{+}_{\boldsymbol{q}}$ and $-\omega^{-}_{\boldsymbol{q}}$ in Eq.~\eqref{eqn13}, and of the sublattice magnetizations in Eq.~\eqref{eqn18}. These expressions allows us to calculate the temperature-dependent mesoscopic parameters via direct comparison to the excitation frequencies of the continuum model. Our theory is numerically validated in Sec.~\ref{sec3}, where the proposed analytical expressions are compared to numerical Monte Carlo simulations for a specific spin model, where the excitation frequencies can be given in a simpler form.

\subsection{Effective temperature-dependent parameters\label{sec2b}}

In the long-wavelength limit, the Hamiltonian in Eq.~\eqref{eqn1} can be approximated by the free-energy functional
\begin{eqnarray}
\mathcal{F}&=&\int\frac{1}{2}\sum_{r,s}\left(\sum_{\alpha,\beta,\gamma}\mathcal{J}_{m}^{rs,\alpha\beta}\partial_{\alpha}m_{r}^{\gamma}\partial_{\beta}m_{s}^{\gamma}+\sum_{\alpha,\beta}\mathcal{D}_{m}^{rs,\alpha\beta}L^{rs,\alpha\beta}\right)\nonumber
\\
&&-\sum_{r,s}\mathcal{K}_{m}^{rs}m_{r}^{z}m_{s}^{z}-\sum_{r}M_{r}B^{z}m_{r}^{z}-\mathcal{J}_{m0}^{AB}\boldsymbol{m}_{A}\boldsymbol{m}_{B}
\textrm{d}^{d}\boldsymbol{r}\label{eqn20}
\end{eqnarray}
in $d$ spatial dimensions.
Here, the magnetization fields $\boldsymbol{m}_{r}$ are required to be of unit length, being defined as $M_{r}m_{r}^{z}=\mu_{r}\left<\tilde{S}_{r}^{z}\right>/V_{\textrm{c}}$, where $V_{\textrm{c}}$ is the volume of the unit cell and $M_{r}$ is the saturation magnetization of the sublattice. The $m$ subscript denotes effective mesoscopic parameters, while $\alpha,\beta$ and $\gamma$ are Cartesian indices. The first line of Eq.~\eqref{eqn20} describes energy contributions from a spatially inhomogeneous magnetization. %For simplicity of the notation, the exchange stiffness $\mathcal{J}_{m}^{rs}$ is assumed to be isotropic not only in spin space but also in real space as denoted by the partial derivatives $\partial_{\alpha}$; this is satisfied in, e.g., hypercubic lattices. The Dzyaloshinsky--Moriya vector is assumed to be pointing along the $z$ direction as in the atomistic model, and it is assumed to be finite between spins which are separated along the $x$ axis. This functional form of the Dzyaloshinsky-Moriya interaction is appropriate for the model system described in Sec.~\ref{sec3}, while it has to be substituted by the appropriate 
In the Dzyaloshinsky--Moriya term, the Lifshitz invariant $L^{rs,\alpha\beta}=1/2\sum_{\gamma,\delta}\varepsilon^{\alpha\gamma\delta}\left(m_{r}^{\gamma}\partial_{\beta}m_{s}^{\delta}-m_{r}^{\delta}\partial_{\beta}m_{s}^{\gamma}\right)$ depends on the considered symmetry class~\cite{Schweflinghaus2014}. The second line of Eq.~\eqref{eqn20} remains finite for homogeneous sublattice magnetizations, describing the energy contribution depending on the global orientation of the magnetization vectors with respect to the easy axis, to the external field, and to each other in the two sublattices.

The equation of motion in the continuum model reads
\begin{eqnarray}
\partial_{t}\boldsymbol{m}_{r}=-\gamma \boldsymbol{m}_{r}\times\left(-\frac{1}{M_{r}}\frac{\delta \mathcal{F}}{\delta \boldsymbol{m}_{r}}\right),\label{eqn20a}
\end{eqnarray}
which transforms into a form analogous to Eq.~\eqref{eqn1e} in the local coordinate system and in Fourier space. 
Requiring that the excitation frequencies of the continuum model coincide with Eq.~\eqref{eqn13} of the atomistic model in the long-wavelength limit, the following temperature dependence is obtained for the parameters:
\begin{align}
\mathcal{J}_{m}^{rs,\alpha\beta}=&\frac{1}{2V_{\textrm{c}}}\sum_{j}\left[J_{ij}^{rs}+\alpha_{0}\left(J_{ij}^{rs}+\Delta J_{ij}^{rs}\right)\textrm{Re}\left<\tilde{S}_{js}^{(1)}\tilde{S}_{ir}^{(2)}\right>\right]\nonumber
\\
&\times R_{ij}^{\alpha}R_{ij}^{\beta}\left<\tilde{S}_{r}^{z}\right>\left<\tilde{S}_{s}^{z}\right>,\label{eqn21}
\\
\mathcal{D}_{m}^{rs,z\beta}=&-\frac{1}{V_{\textrm{c}}}\sum_{j}\left[D_{ij}^{z,rs}+\alpha_{0}\left(J_{ij}^{rs}+\Delta J_{ij}^{rs}\right)\textrm{Im}\left<\tilde{S}_{js}^{(1)}\tilde{S}_{ir}^{(2)}\right>\right]\nonumber
\\
&\times R_{ij}^{\beta}\left<\tilde{S}_{r}^{z}\right>\left<\tilde{S}_{s}^{z}\right>,\label{eqn22}
\\
\mathcal{K}_{m}^{rs}=&\frac{1}{V_{\textrm{c}}}\left[K^{r}\delta_{rs}\left(1-\alpha_{0}\left<\tilde{S}_{ir}^{(1)}\tilde{S}_{ir}^{(2)}\right>\right)+\frac{1}{2}\sum_{j}\Delta J_{ij}^{rs}\left(1-\alpha_{0}\right.\right.\nonumber
\\
&\left.\left.\times\textrm{Re}\left<\tilde{S}_{js}^{(1)}\tilde{S}_{ir}^{(2)}\right>\right)+\alpha_{0}D_{ij}^{rs}\textrm{Im}\left<\tilde{S}_{js}^{(1)}\tilde{S}_{ir}^{(2)}\right>\right]\left<\tilde{S}_{r}^{z}\right>\left<\tilde{S}_{s}^{z}\right>,\label{eqn23}
\\
\mathcal{J}_{m0}^{AB}=&\frac{1}{V_{\textrm{c}}}\sum_{j}\left(J_{ij}^{AB}+\alpha_{0}\left(J_{ij}^{AB}+\Delta J_{ij}^{AB}\right)\textrm{Re}\left<\tilde{S}_{jB}^{-}\tilde{S}_{iA}^{-}\right>\right)\nonumber
\\
&\times\left<\tilde{S}_{A}^{z}\right>\left<\tilde{S}_{B}^{z}\right>,\label{eqn24}
\end{align}
where $\boldsymbol{R}_{ij}=\boldsymbol{R}_{i}-\boldsymbol{R}_{j}$. The comparison based on the magnon spectrum only provides information on the $z$ component of the Dzyaloshinsky--Moriya interaction; the other components may be obtained by comparing the atomistic and continuum models for ground states oriented along different directions.
%is the distance between the sites $i$ and $j$ along the $x$ axis.

The parameters Eq.~\eqref{eqn21}-\eqref{eqn24} show similar trends to what has been calculated in single-sublattice ferromagnetic systems in Refs.~\cite{Bastardis,Rozsa,Evans}. The main contribution to the temperature dependence of all the parameters comes from $\left<\tilde{S}_{r}^{z}\right>\left<\tilde{S}_{s}^{z}\right>$, which corresponds to the mean-field or random-phase approximations. By considering a decoupling scheme different from the random-phase approximation, i.e., $\alpha_0\neq 0$, the temperature dependence of the micromagnetic parameters is corrected by taking spin correlation effects into account. The correlation corrections  proportional to $\alpha_{0}$ have a positive sign for the isotropic exchange and the Dzyaloshinsky--Moriya terms, which makes these terms decrease slower in magnitude with the temperature. 
Note that for only nearest-neighbor interactions, the relative correction to the isotropic exchange and Dzyaloshinsky--Moriya interactions turn out to be precisely the same, similar to what is observed in a single-sublattice ferromagnet in Ref.~\cite{Rozsa}.
In contrast, the correlation corrections are negative for the anisotropy terms, indicating a faster decrease. 
In ferromagnets, this is known to correspond to the Callen--Callen law $\mathcal{K}\sim \langle S^{z}\rangle^3$ for the temperature dependence of the uniaxial anisotropy~\cite{Callen2} based on the first term in Eq.~\eqref{eqn23}, and to a scaling exponent $\mathcal{K}\sim \langle S^{z}\rangle^{2+\varepsilon}$ slightly larger than $2$ for the two-ion anisotropy~\cite{Evans} in the second term. 
The last term in Eq.~\eqref{eqn23} gives a positive contribution to the anisotropy, meaning that the Dzyaloshinsky--Moriya interaction stabilizes collinear order in the presence of thermal fluctuations~\cite{Rozsa}. 

Equation~\eqref{eqn21} defines different mesoscopic exchange stiffness parameters for the intrasublattice coupling $\mathcal{J}_{m}^{AA},\mathcal{J}_{m}^{BB}$ and for the intersublattice coupling $\mathcal{J}_{m}^{AB},\mathcal{J}_{m}^{BA}$. 
Together with the anisotropy, the value of these exchange stiffness parameters are relevant for the estimation of domain wall width $\delta_w (T) \propto \sqrt{\mathcal{K}(T)/\mathcal{J}(T)}$. Equation~\eqref{eqn22} describes the temperature dependence of the mesoscopic intrasublattice and intersublattice Dzyaloshinsky-Moriya parameters, which are necessary for the estimation of the skyrmion radius~\cite{Barker2016,Tomasello2018}. The competition between the different contributions to the anisotropy term in Eq.~\eqref{eqn23} gives rise to fluctuation-driven spin reorientation transitions induced by the Dzyaloshinsky--Moriya interaction~\cite{Nagyfalusi} and unusual exponents in the temperature dependence of the anisotropy parameter~\cite{Evans}, similarly to what has been observed before in ferromagnetic systems. 

As mentioned in Sec.~\ref{sec2a}, in the quantum case $\tilde{S}_{js}^{(1)}\tilde{S}_{ir}^{(2)}$ has to be replaced by the anticommutator $1/2\left[\tilde{S}_{js}^{(1)},\tilde{S}_{ir}^{(2)}\right]_{+}$. For $S=1/2$, this choice enforces the coefficient of the single-ion anisotropy $K^{r}$ to vanish ($\alpha_{0}=1/\left(2S^{2}\right)$ and $1/2\left[\tilde{S}_{ir}^{(1)},\tilde{S}_{ir}^{(2)}\right]_{+}=1/4$), which is consistent with the fact that the single-ion anisotropy just acts as a constant energy term. Indeed, this condition was one of the motivations behind choosing the value of the decoupling coefficient $\alpha_{0}$ in Ref.~\cite{Anderson}.

\subsection{Scaling exponents\label{sec2c}}

To obtain a simpler formula for the temperature dependence of the parameters in the continuum model, we only keep the Heisenberg interactions which are typically the largest in magnitude, and calculate the expressions
\begin{align}
\mathcal{J}_{m0}^{rs}=&\frac{1}{V_{\textrm{c}}}\sum_{j}\left(J_{ij}^{rs}+\alpha_{0}\left(J_{ij}^{rs}+\Delta J_{ij}^{rs}\right)\textrm{Re}\left<\tilde{S}_{js}^{(1)}\tilde{S}_{ir}^{(2)}\right>\right)\nonumber
\\
&\times\left<\tilde{S}_{r}^{z}\right>\left<\tilde{S}_{s}^{z}\right>,\label{eqn24a}
\end{align}
which are connected to Eqs.~\eqref{eqn21} and \eqref{eqn24}. The correlation functions are real in the absence of the Dzyaloshinsky--Moriya interaction, and we substitute them into Eq.~\eqref{eqn24a} from Eq.~\eqref{eqn1h} using Eq.~\eqref{eqn1j}, and approximate $\underline{\underline{\tilde{\Gamma}}}_{\boldsymbol{q}}$ with the spin-wave Hamiltonian $\underline{\underline{\tilde{H}}}_{\textrm{SW},\boldsymbol{q}}$ from Eq.~\eqref{eqn1f} in the low-temperature limit. This results in
\begin{align}
\underline{\underline{\mathcal{J}}}_{m0}=&\frac{1}{V_{\textrm{c}}}\left(\underline{\underline{\mathfrak{J}}}_{\boldsymbol{0}}+\frac{2\alpha_{0}k_{\textrm{B}}T}{N_{\textrm{c}}}\sum_{q}\underline{\underline{\mathfrak{J}}}_{\boldsymbol{0}}\circ\left[\underline{\underline{\sigma}}^{z}\left(\underline{\underline{\tilde{H}}}_{\textrm{SW},\boldsymbol{q}}^{-1}\right)^{T}\right]\right)\nonumber
\\
&\circ\left[\left<\underline{\tilde{S}}^{z}\right>\left<\left(\underline{\tilde{S}}^{z}\right)^{T}\right>\right].\label{eqn24b}
\end{align}

As mentioned above, the temperature dependence of the parameters of the continuum model is often expressed in terms of a power law of the magnetization. This is common practice partially because it is easier to implement numerically in a micromagnetic framework and partially because in non-equilibrium situations the value of the magnetization represents better the thermodynamic state of the system than the temperature of the heat bath. Both antiferromagnetic and ferrimagnetic systems may be characterized by the sublattice magnetizations $\left<\tilde{S}_{A}^{z}\right>,\left<\tilde{S}_{B}^{z}\right>$. As it is shown in Eqs.~\eqref{eqn21}-\eqref{eqn24b}, the sublattice magnetizations are a more natural choice for expressing the effective parameters than the combinations $\mu_{A}\left<\tilde{S}_{A}^{z}\right>\pm\mu_{B}\left<\tilde{S}_{B}^{z}\right>$ resulting in the total and staggered magnetizations, respectively. The temperature $T$ in Eq.~\eqref{eqn24b} may also be expressed by either sublattice magnetization using the low-temperature expansion of Eq.~\eqref{eqn18},
\begin{align}
1-\left<\tilde{S}_{r}^{z}\right>\approx\Phi_{r}\approx\frac{k_{\textrm{B}}T}{N_{\textrm{c}}}\sum_{q}\left[\underline{\underline{\sigma}}^{z}\left(\underline{\underline{\tilde{H}}}_{\textrm{SW},\boldsymbol{q}}^{-1}\right)^{T}\right]^{rr}.\label{eqn24c}
\end{align}
Equation~\eqref{eqn24c} connects the two sublattice magnetizations to each other as well, meaning that either one can be used to express the effective parameters. To simplify the expressions further, we go to the classical limit, assume that all intrasublattice interaction terms are the same and there is no external magnetic field. In this case, it can be shown based on the definition Eq.~\eqref{eqn18} that $\left<\tilde{S}_{A}^{z}\right>=\left<\tilde{S}_{B}^{z}\right>$, meaning that the temperature dependence of all magnetizations is precisely the same if they are normalized to their zero-temperature value.
This is the case for antiferromagnets with identical sublattices, but is also a good approximation for ferrimagnets with $\mu_{A}\neq\mu_{B}$ if the intrasublattice interactions are negligible compared to the intersublattice ones. This follows from the fact that the self-consistency Eqs.~\eqref{eqn1g}, \eqref{eqn1j} and \eqref{eqn18} do not depend explicitly on the magnetic moments. In the following we restrict our attention to this limit, and leave the case of different sublattice magnetizations observable in, e.g., ferrimagnets with a compensation point or in the quantum limit, to later studies.

%If all terms in the Hamiltonian Eq.~\eqref{eqn1} are negligible compared to the intersublattice isotropic exchange $J_{ij}^{AB}$, using Eqs.~\eqref{eqn4}-\eqref{eqn7}, \eqref{eqn14}, \eqref{eqn18}, and \eqref{eqn21}, in the low-temperature limit the effective exchange interaction may be expressed as
As a specific example, we consider nearest-neighbor antiferromagnetic exchange $\mathfrak{J}_{\boldsymbol{0}}^{AB}=\mathfrak{J}_{\boldsymbol{0}}^{BA}=-\left(1-\lambda\right)\mathfrak{J}_{\boldsymbol{0}}$ and $\mathfrak{J}_{\boldsymbol{0}}^{AA}=\mathfrak{J}_{\boldsymbol{0}}^{BB}=\lambda\mathfrak{J}_{\boldsymbol{0}}$, where $\mathfrak{J}_{\boldsymbol{0}}$ determines the absolute strength of the interactions and $\lambda\in\left[0,1\right]$ is a scaling parameter. Changing $\lambda$ transforms from a nearest-neighbor antiferromagnetic model to two decoupled ferromagnetic sublattices, while keeping the mean-field critical temperature $k_{\textrm{B}}\Theta=\left(\mathfrak{J}_{\boldsymbol{0}}^{AA}-\mathfrak{J}_{\boldsymbol{0}}^{AB}\right)/3=\mathfrak{J}_{\boldsymbol{0}}/3$ constant. Substituting Eq.~\eqref{eqn24c} into Eq.~\eqref{eqn24b} and using Callen's decoupling with $2\alpha_{0}=1$ yields
\begin{eqnarray}
\mathcal{J}_{m0}^{rs}\propto\left<\tilde{S}^{z}\right>^{2}\left[1+\varepsilon^{rs}\left(1-\left<\tilde{S}^{z}\right>\right)\right]\approx\left<\tilde{S}^{z}\right>^{2-\varepsilon^{rs}},\label{eqn25}
\end{eqnarray}
where $\left<\tilde{S}^{z}\right>$ is the magnetization on either sublattice and the correction to the intersublattice ($AB,BA$) and intrasublattice ($AA,BB$) exponents read
\begin{align}
\varepsilon^{AB}=\varepsilon^{BA}=\varepsilon_{\textrm{inter}}=&\frac{\frac{1}{N_{\textrm{c}}}\sum_{\boldsymbol{q}}\frac{\left(1-\lambda\right)\left(\gamma_{\boldsymbol{q}}^{\textrm{inter}}\right)^{2}}{\left(1-\lambda\gamma_{\boldsymbol{q}}^{\textrm{intra}}\right)^{2}-\left(1-\lambda\right)^{2}\left(\gamma_{\boldsymbol{q}}^{\textrm{inter}}\right)^{2}}}{\frac{1}{N_{\textrm{c}}}\sum_{\boldsymbol{q}}\frac{1-\lambda\gamma_{\boldsymbol{q}}^{\textrm{intra}}}{\left(1-\lambda\gamma_{\boldsymbol{q}}^{\textrm{intra}}\right)^{2}-\left(1-\lambda\right)^{2}\left(\gamma_{\boldsymbol{q}}^{\textrm{inter}}\right)^{2}}},\label{eqn26a}
\\
\varepsilon^{AA}=\varepsilon^{BB}=\varepsilon_{\textrm{intra}}=&\frac{\frac{1}{N_{\textrm{c}}}\sum_{\boldsymbol{q}}\frac{\gamma_{\boldsymbol{q}}^{\textrm{intra}}\left(1-\lambda\gamma_{\boldsymbol{q}}^{\textrm{intra}}\right)}{\left(1-\lambda\gamma_{\boldsymbol{q}}^{\textrm{intra}}\right)^{2}-\left(1-\lambda\right)^{2}\left(\gamma_{\boldsymbol{q}}^{\textrm{inter}}\right)^{2}}}{\frac{1}{N_{\textrm{c}}}\sum_{\boldsymbol{q}}\frac{1-\lambda\gamma_{\boldsymbol{q}}^{\textrm{intra}}}{\left(1-\lambda\gamma_{\boldsymbol{q}}^{\textrm{intra}}\right)^{2}-\left(1-\lambda\right)^{2}\left(\gamma_{\boldsymbol{q}}^{\textrm{inter}}\right)^{2}}},\label{eqn26}
\end{align}
with the geometrical factors
\begin{eqnarray}
\gamma_{\boldsymbol{q}}^{rs}\sum_{\boldsymbol{R}_{i}-\boldsymbol{R}_{j}}J_{ij}^{rs}=\sum_{\boldsymbol{R}_{i}-\boldsymbol{R}_{j}}\textrm{e}^{-\textrm{i}\boldsymbol{q}\left(\boldsymbol{R}_{i}-\boldsymbol{R}_{j}\right)}J_{ij}^{rs}.\label{eqn27}
\end{eqnarray}

\begin{figure}
    \centering
    \includegraphics[width = \columnwidth]{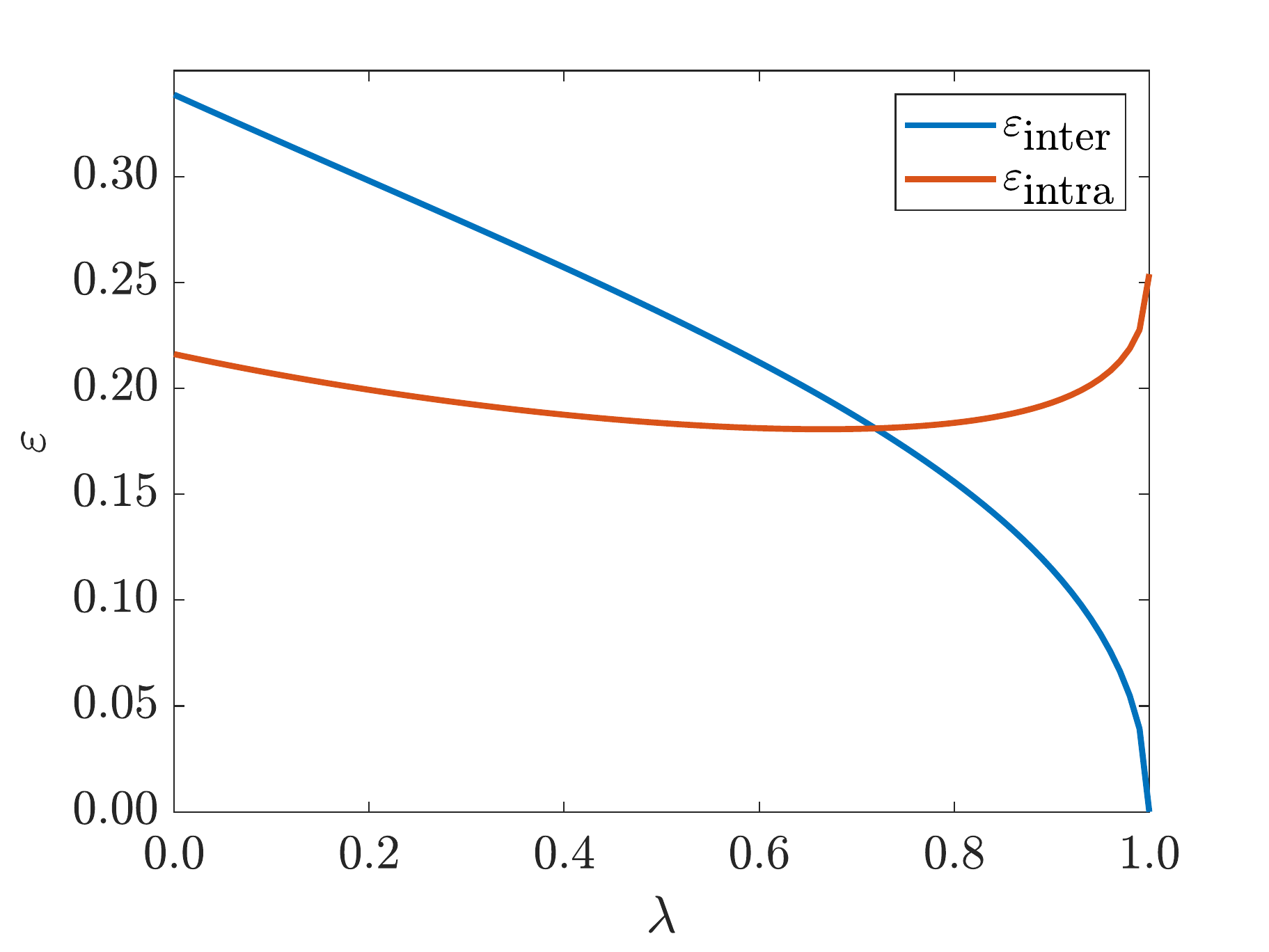}
    \caption{Corrections to the scaling exponents in the rock-salt structure based on Eqs. \eqref{eqn26a} and \eqref{eqn26}. The inter- and intersublattice Heisenberg interactions are scaled from only antiferromagnetic coupling between the sublattices $\lambda=0$ to two decoupled ferromagnetic sublattices $\lambda=1$.}\label{fig0}
\end{figure}

The $\varepsilon$ values from Eqs. \eqref{eqn26a} and \eqref{eqn26} are shown in Fig.~\ref{fig0} for the rock-salt structure, where the antiferromagnetically coupled sublattices together form a simple cubic lattice and each sublattice is an fcc lattice. For $\lambda=1$ (ferromagnet), we numerically obtain an exponent close to the analytical value $\varepsilon_{\textrm{intra}}=0.255$, which was reported for the ferromagnetic fcc lattice in Ref.~\cite{Atxitia2010}. The intersublattice correction to the exponent vanishes in this limit as can be seen from Eq.~\eqref{eqn26a}, meaning that for weak intersublattice coupling parameters, the corresponding effective parameter closely follows a mean-field scaling. In the $\lambda=0$ case (antiferromagnet), the intersublattice exponent also converges to the analytical value for a simple cubic lattice $\varepsilon^{\textrm{sc}}=0.341$, although the coupling between the sublattices is antiferromagnetic instead of ferromagnetic as in Ref.~\cite{Atxitia2010}. The correction to the intrasublattice exponent remains finite in this case, meaning that the influence of spin correlations on the temperature dependence of intrasublattice coupling can be observed even in this limit. Increasing $\lambda$ leads to a decrease in both the intersublattice and intrasublattice $\varepsilon$ values at first, because distributing the exchange interactions between nearest and next-nearest neighbors brings the system closer to a mean-field behavior. It was similarly found in Ref.~\cite{Atxitia2010} that the $\varepsilon$ value is lower for ferromagnetic FePt where interactions with several neighbors were taken into account than for any nearest-neighbor cubic lattice. However, the scaling performed here demonstrates that this decreasing trend is reversed for $\lambda$ values close to $1$ in the intrasublattice term, while the intersublattice $\varepsilon$ value vanishes as discussed above.

%The power law in Eq.~\eqref{eqn25} is not only similar to the ferromagnetic case discussed in, e.g., Ref.~\cite{Bastardis}, but also numerically identical to it.
The agreement for the exponent corrections between the ferromagnetic and the antiferromagnetic case mentioned for $\lambda=0$ in the simple cubic lattice is a general property of the model. If it is assumed that the antiferromagnetic alignment of the spins is realized in a system where all atoms together form a Bravais lattice, then one obtains $\gamma_{\boldsymbol{q}+\boldsymbol{Q}}^{AB}=-\gamma_{\boldsymbol{q}}^{AB}$ with $\boldsymbol{Q}$ the wave vector of the antiferromagnetic ordering, making it possible to rewrite Eq.~\eqref{eqn26a} for $\lambda=0$ as %the correction to the exponent as
\begin{eqnarray}
\varepsilon_{\textrm{inter}}=\frac{\frac{1}{2N_{\textrm{c}}}\sum_{\boldsymbol{q},\textrm{FM}}\frac{\gamma_{\boldsymbol{q}}^{\textrm{inter}}}{1-\gamma_{\boldsymbol{q}}^{\textrm{inter}}}}{\frac{1}{2N_{\textrm{c}}}\sum_{\boldsymbol{q},\textrm{FM}}\frac{1}{1-\gamma_{\boldsymbol{q}}^{\textrm{inter}}}},\label{eqn28}
\end{eqnarray}
where the summation now runs over the atomic or ferromagnetic Brillouin zone which is twice the size of the antiferromagnetic one. For infinite lattices where the summations can be replaced by integrals, the correction to the exponent is $\varepsilon_{\textrm{inter}}^{\textrm{sc}}=0.341$ for the simple cubic and $\varepsilon_{\textrm{inter}}^{\textrm{bcc}}=0.282$ for the body-centered cubic lattice~\cite{Atxitia2010}, both of which can accommodate a two-sublattice ordering. Even for systems where all atoms together do not form a Bravais lattice (e.g., the honeycomb lattice), it can be derived that the expectation values and the correlation functions in the antiferromagnetically aligned model precisely coincide with those of the ferromagnetic model where the sign of all intersublattice coupling terms is reversed. Consequently, Eqs.~\eqref{eqn26a} and \eqref{eqn26} may also be used for ferromagnets containing both nearest-neighbor and next-nearest-neighbor interactions. The agreement between the ferromagnetic and antiferromagnetic cases essentially relies on the fact that in the classical limit, %although not in the quantum case, 
the self-consistency Eqs.~\eqref{eqn1g}, \eqref{eqn1j} and \eqref{eqn18} %Eqs.~\eqref{eqn4}-\eqref{eqn7} and \eqref{eqn14}-\eqref{eqn17} 
do not depend on the magnon frequencies in Eq.~\eqref{eqn13}, which are different between the ferromagnetic and the antiferromagnetic alignment. This is different in the quantum case, where Eq.~\eqref{eqn1i} does depend on the frequencies as shown in Appendix~\ref{appendixA}.

For weak Dzyaloshinsky--Moriya interaction, the same correction $\varepsilon$ to the scaling exponent can be used. For two-site anisotropy between the same pairs of atoms as the exchange, the exponent is $2+\varepsilon$ owing to the opposite sign of the correlation correction between Eqs.~\eqref{eqn21}-\eqref{eqn22} and the second term in Eq.~\eqref{eqn23}, respectively. For the on-site anisotropy term, the exponent is close to $3$ as in the ferromagnetic case~\cite{Callen2} since the on-site correlations are stronger than the two-site terms.

In two-dimensional systems, the sums in Eqs.~\eqref{eqn26a} and \eqref{eqn26} diverge for infinite lattice sizes, as is known from, e.g., the proof of the Mermin--Wagner theorem~\cite{Mermin}. This implies that the exponents may only be calculated if a finite anisotropy is taken into account, in which case they have to be evaluated numerically. Since the correlation corrections are expected to be enhanced in low-dimensional systems, this procedure is carried out and compared to numerical simulations in Sec.~\ref{sec3}.

\section{Simulations\label{sec3}}

To probe the accuracy of the analytical method described in Sec.~\ref{sec2}, its predictions will be compared to the numerical simulations based on the Hamiltonian Eq.~\eqref{eqn1}. While the magnetization and the static correlation functions may be directly determined from averaging over spin configurations from the different simulation steps, the frequencies required for determining the temperature dependence of the parameters in the continuum model are more difficult to access. Equations~\eqref{eqn1h}-\eqref{eqn1j} establish the relations between the expectation values and the frequencies. They may be reformulated as
\begin{eqnarray}
&&\left<S_{-\boldsymbol{q}A}^{+}S_{\boldsymbol{q}A}^{-}\right>\left<S_{-\boldsymbol{q}B}^{+}S_{\boldsymbol{q}B}^{-}\right>-\left<S_{-\boldsymbol{q}B}^{+}S_{\boldsymbol{q}A}^{-}\right>\left<S_{-\boldsymbol{q}A}^{+}S_{\boldsymbol{q}B}^{-}\right>\nonumber
\\
&&=4\left<S_{A}^{z}\right>\left<S_{B}^{z}\right>\frac{1}{N_{\textrm{c}}^{2}}\frac{\gamma}{\mu_{A}}\frac{\gamma}{\mu_{B}}\frac{\left(k_{\textrm{B}}T\right)^{2}}{\omega^{+}_{\boldsymbol{q}}\omega^{-}_{\boldsymbol{q}}},\label{eqn29}
\\
&&\frac{\left<S_{-\boldsymbol{q}A}^{+}S_{\boldsymbol{q}A}^{-}\right>}{2\gamma\mu_{A}^{-1}\left<S_{A}^{z}\right>}+\frac{\left<S_{-\boldsymbol{q}B}^{+}S_{\boldsymbol{q}B}^{-}\right>}{2\gamma\mu_{B}^{-1}\left<S_{B}^{z}\right>}=\frac{1}{N_{\textrm{c}}}\frac{k_{\textrm{B}}T\left(\omega^{+}_{\boldsymbol{q}}+\omega^{-}_{\boldsymbol{q}}\right)}{\omega^{+}_{\boldsymbol{q}}\omega^{-}_{\boldsymbol{q}}}.\label{eqn30}
\end{eqnarray}
The product of the frequencies $\omega_{\textrm{prod}}=\omega^{+}_{\boldsymbol{q}}\omega^{-}_{\boldsymbol{q}}$ is given by Eq.~\eqref{eqn29}, which also yields the sum $\omega_{\textrm{sum}}=\omega^{+}_{\boldsymbol{q}}+\omega^{-}_{\boldsymbol{q}}$ from Eq.~\eqref{eqn30}.
The individual frequencies may be calculated as
\begin{eqnarray}
\omega^{\pm}_{\boldsymbol{q}}=\frac{1}{2}\left[\omega_{\textrm{sum}}\pm\sqrt{\omega_{\textrm{sum}}^{2}-4\omega_{\textrm{prod}}}\right].\label{eqn31}
\end{eqnarray}
Note that in Eqs.~\eqref{eqn29} and \eqref{eqn30}, the correlation functions are given in the global coordinate system for easier implementation in the simulations. Since these equations establish a connection between the eigenfrequencies, the correlation functions and the temperature, they may be considered as a form of the equipartition theorem. Although Eqs.~\eqref{eqn29} and \eqref{eqn30} were determined from the Green's function formalism, they do not depend on the explicit form of the decoupling $\alpha_{0}$, only on the assumption that the spectral density is concentrated in single-particle excitations. Therefore, substituting the expectation values obtained from the simulations into Eq.~\eqref{eqn29} and \eqref{eqn30} enables the calculation of the frequencies of the simulated system. Furthermore, this method allows for determining the frequencies based on Monte Carlo simulations, which accurately describe thermal equilibrium properties but do not provide direct access to the real-time dynamics of the system.

\begin{figure}
    \centering
    \includegraphics[width = \columnwidth]{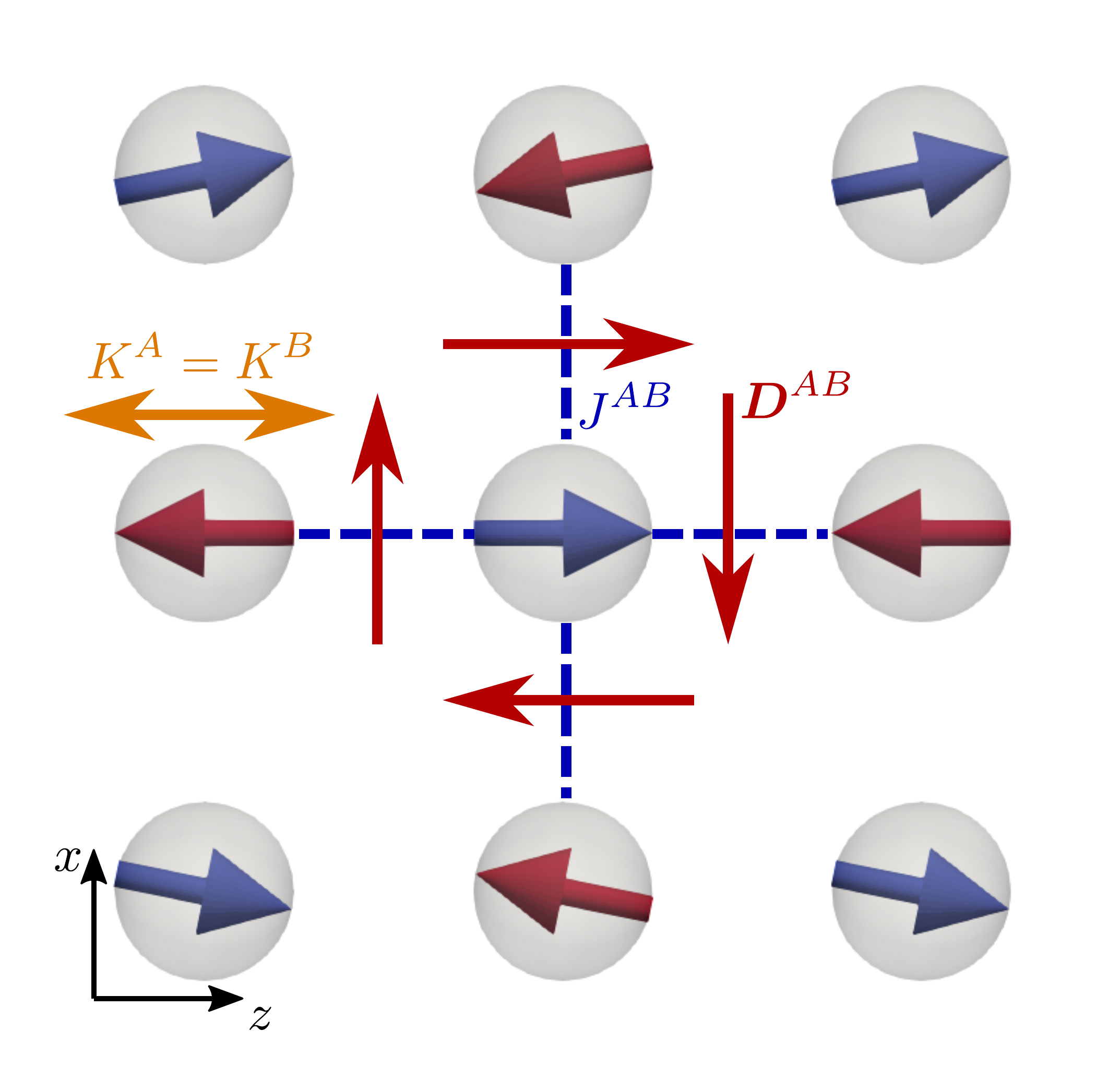}
    \caption{Sketch of the system used for the simulations. The spin directions illustrate a spin wave propagating along the $x$ direction on an antiferromagnetic background along the $z$ direction.
    $K^A=K^B=K$ stand for the uniaxial anisotropy constants, $-J^{AB}=J$ the intersublattice antiferromagnetic exchange parameter and $D^{AB}=D$ the Dzyaloshinsky--Moriya interaction parameter.}\label{fig1}
\end{figure}

The simulated model system is illustrated in Fig.~\ref{fig1}. It consists of a square lattice with equivalent sublattices $\mu_{A}=\mu_{B}=\mu_{\textrm{S}}$, only considering nearest-neighbor intersublattice Heisenberg exchange $-J^{AB}=J>0$ and Dzyaloshinsky--Moriya interactions of magnitude $D^{AB}=D$, with the Dzyaloshinsky--Moriya vectors being perpendicular to the lattice vectors connecting the neighbours following a $C_{4\textrm{v}}$ symmetry. The easy axis $K^{A}=K^{B}=K$ was assumed to lie along one of the nearest-neighbour directions, which enables the investigation of the Dzyaloshinsky--Moriya vectors parallel to the $z$ direction on the spin-wave spectrum. The external magnetic field was set to zero. We performed Monte Carlo simulations on a $64\times 64$ lattice using the single-spin Metropolis algorithm where the trial spin direction is chosen uniformly on the surface of the unit sphere. The lattice was equilibrated for $2\cdot10^{5}$ Monte Carlo steps at each temperature, then the expectation values were calculated from data obtained over $10^{8}$ Monte Carlo steps. To further improve the accuracy, $50$ independent simulations were averaged in the end.

\begin{figure}
    \centering
    \includegraphics[width = \columnwidth]{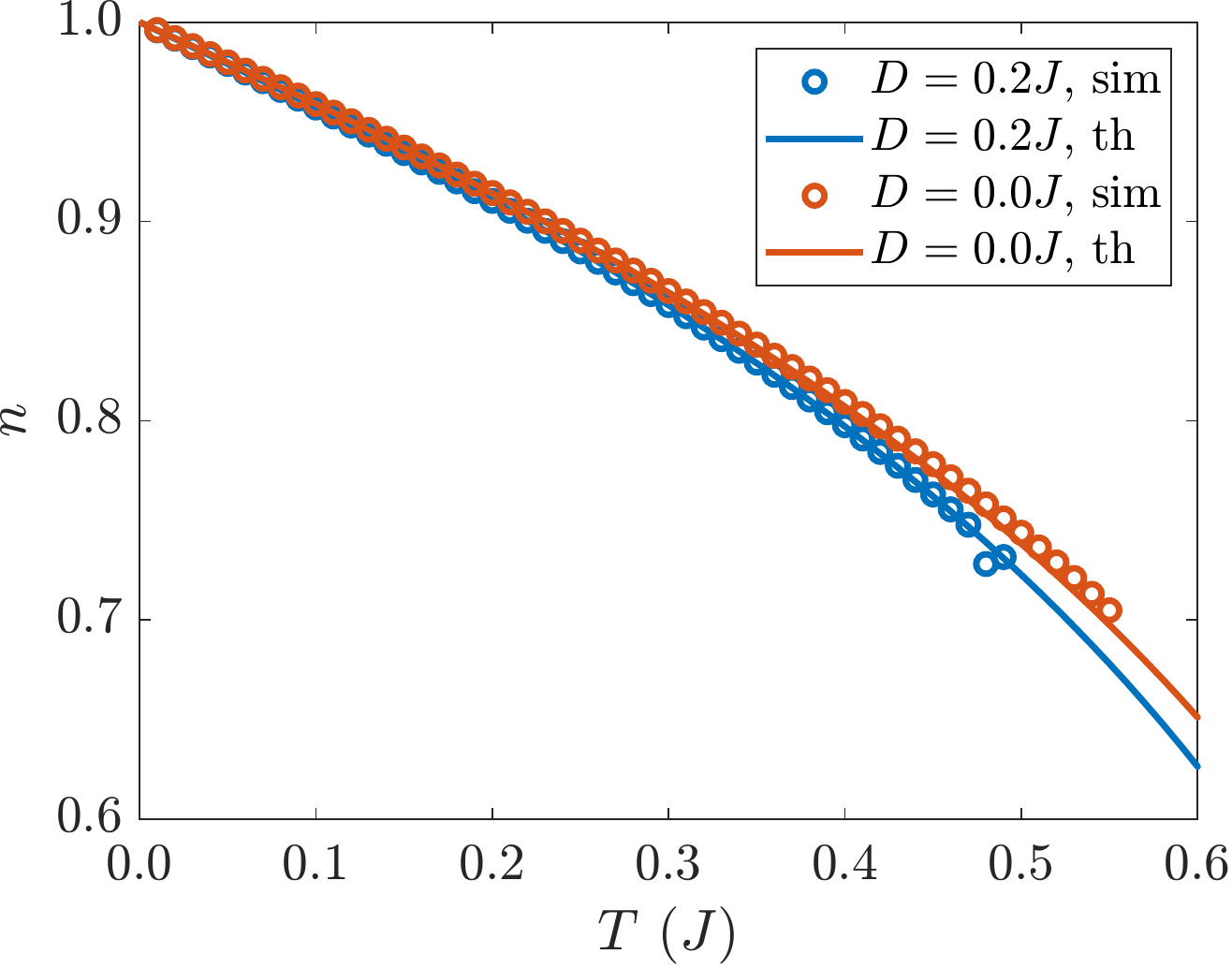}
    \caption{Temperature dependence of the staggered magnetization. Results of the numerical simulations (symbols) are compared to Green's function theory calculations from Eq.~\eqref{eqn18} (lines). The atomistic model parameters are $D=0.2J$ for the blue curves and $D=0.0J$ for the orange curves, and the anisotropy is $K=0.1J$.\label{fig2a}}
\end{figure}

Due to the symmetry of the sublattices, we obtain $\left<\tilde{S}^{z}_{A}\right>=\left<\tilde{S}^{z}_{B}\right>=\left<S^{z}\right>$, which also coincides with the dimensionless staggered magnetization $n$. The simulated and calculated values of $n$ are compared in Fig.~\ref{fig2a}, demonstrating good agreement. Including the Dzyaloshinsky--Moriya interaction decreases the staggered magnetization at a fixed temperature. The critical temperature of the system is around %$k_{\textrm{B}}T_{\textrm{c}}\approx 0.75J$ based on the squared magnetization in the simulations
$k_{\textrm{B}}T_{\textrm{c}}\approx 0.84J$ from Green's function theory. Note that $\left<S^{z}\right>$ is not possible to calculate accurately at temperatures close to $T_{\textrm{c}}$, since due to the relatively small system size and the long simulation length the system starts to switch between the $+z$ and $-z$ directions.

\begin{figure}
    \centering
    \includegraphics[width = \columnwidth]{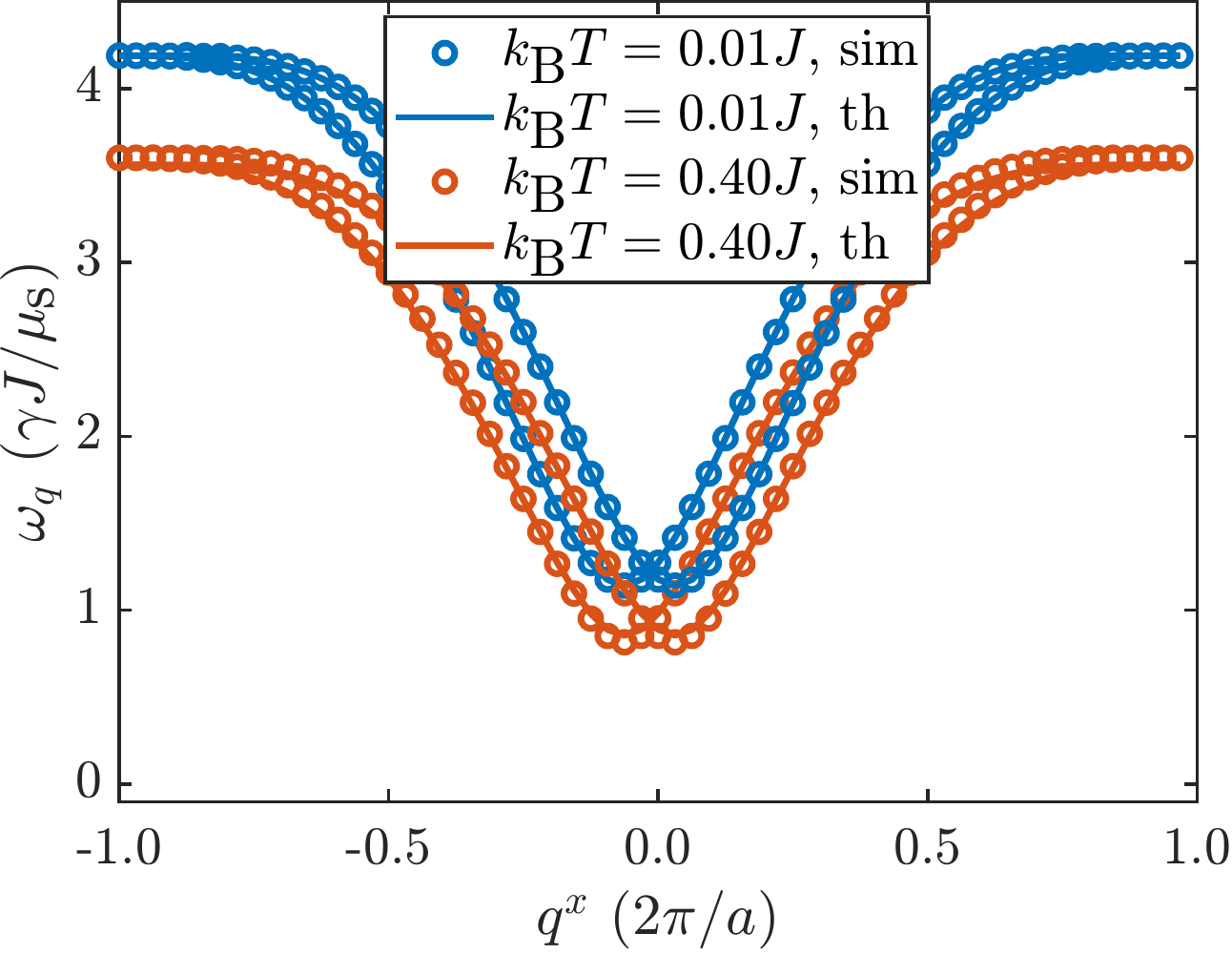}
    \caption{Spin-wave spectrum for $D=0.2J$ and $K=0.1J$. Results of the numerical simulations from Eq.~\eqref{eqn29} (symbols) are compared to Green's function theory calculations from Eqs.~\eqref{eqn32} and \eqref{eqn33} (lines) at two different temperatures.\label{fig2}}
\end{figure}

The spin-wave spectrum at finite temperature was calculated based on Eq.~\eqref{eqn29}, since the symmetry of the sublattices implies $\omega^{+}_{\boldsymbol{q}}=-\omega^{-}_{\boldsymbol{q}}$, %$\left<\tilde{S}^{z}_{A}\right>=\left<\tilde{S}^{z}_{B}\right>=\left<S^{z}\right>$ 
and both sides of Eq.~\eqref{eqn30} vanish. For the considered system, the two branches of the spin-wave dispersion relation are given by
\begin{align}
&\omega^{+}_{\boldsymbol{q}}=\left<S^{z}\right>^{-1}\nonumber
\\
&\times\sqrt{\left(4\mathcal{J}+2\mathcal{K}\right)^{2}-\left(2\mathcal{J}\left[\cos\left(q^{x}a\right)+\cos\left(q^{z}a\right)\right]-2\mathcal{D}\sin\left(q^{x}a\right)\right)^{2}},\label{eqn32}
\\
&-\omega^{-}_{-\boldsymbol{q}}=\left<S^{z}\right>^{-1}\nonumber
\\
&\times\sqrt{\left(4\mathcal{J}+2\mathcal{K}\right)^{2}-\left(2\mathcal{J}\left[\cos\left(q^{x}a\right)+\cos\left(q^{z}a\right)\right]+2\mathcal{D}\sin\left(q^{x}a\right)\right)^{2}}.\label{eqn33}
\end{align}
The spectrum is illustrated in Fig.~\ref{fig2}. The Dzyaloshinsky--Moriya interaction lifts the degeneracy of the two branches and shifts the minimum of the spectrum away from $q^{x}=0$. The anisotropy opens a gap in the spectrum, which is exchange enhanced compared to the ferromagnetic case: 
for $K\ll J$, $\omega_{\boldsymbol{0},\textrm{AFM}} \approx \sqrt{2(4\mathcal{J}) (2\mathcal{K})} \gg 2\mathcal{K}=\omega_{\boldsymbol{0},\textrm{FM}}$.
The theoretical curves are given by Eqs.~\eqref{eqn32} and \eqref{eqn33}, where the parameters $\mathcal{J},\mathcal{D}=\lvert\mathcal{D}_{ij}\rvert,$ and $\mathcal{K}$ are defined as 
\begin{align}
\mathcal{J}=&\left[J+\alpha_{0}J\textrm{Re}\left<\tilde{S}_{jB}^{(1)}\tilde{S}_{iA}^{(2)}\right>\right]\left<\tilde{S}^{z}\right>^{2},\label{eqn34}
\\
\mathcal{D}_{ij}=&\left[D_{ij}-\alpha_{0}J\textrm{Im}\left<\tilde{S}_{jB}^{(1)}\tilde{S}_{iA}^{(2)}\right>\right]\left<\tilde{S}^{z}\right>^{2},\label{eqn35}
\\
\mathcal{K}=&\left[K\left(1-\alpha_{0}\left<\tilde{S}_{iA}^{(1)}\tilde{S}_{iA}^{(2)}\right>\right)-\frac{1}{2}\sum_{j}\alpha_{0}D_{ij}\textrm{Im}\left<\tilde{S}_{jB}^{(1)}\tilde{S}_{iA}^{(2)}\right>\right]\left<\tilde{S}^{z}\right>^{2},\label{eqn36}
\end{align}
based on Eqs.~\eqref{eqn21}-\eqref{eqn23} in an atomistic description. Note that the sign changes in Eqs.~\eqref{eqn35} and \eqref{eqn36} compared to Eqs.~\eqref{eqn22} and \eqref{eqn23} appear due to the sign change in $J$ and the antiferromagnetic alignment of the sublattices, respectively.  
Figure~\ref{fig2} supports the high accuracy of the Green's function formalism up to intermediate temperature values of $k_{\textrm{B}}T=0.40J$.%, where the critical temperature of the system is around %$k_{\textrm{B}}T_{\textrm{c}}\approx 0.75J$ based on the squared magnetization in the simulations
%$k_{\textrm{B}}T_{\textrm{c}}\approx 0.84J$ from Green's function theory. Note that the magnon frequencies cannot be determined at temperatures close to $T_{\textrm{c}}$, since due to the relatively small system size and the long simulation length the system starts to switch between the $+z$ and $-z$ directions, making it impossible to calculate $\left<S^{z}\right>$ accurately.

\begin{figure}
    \centering
    \includegraphics[width = \columnwidth]{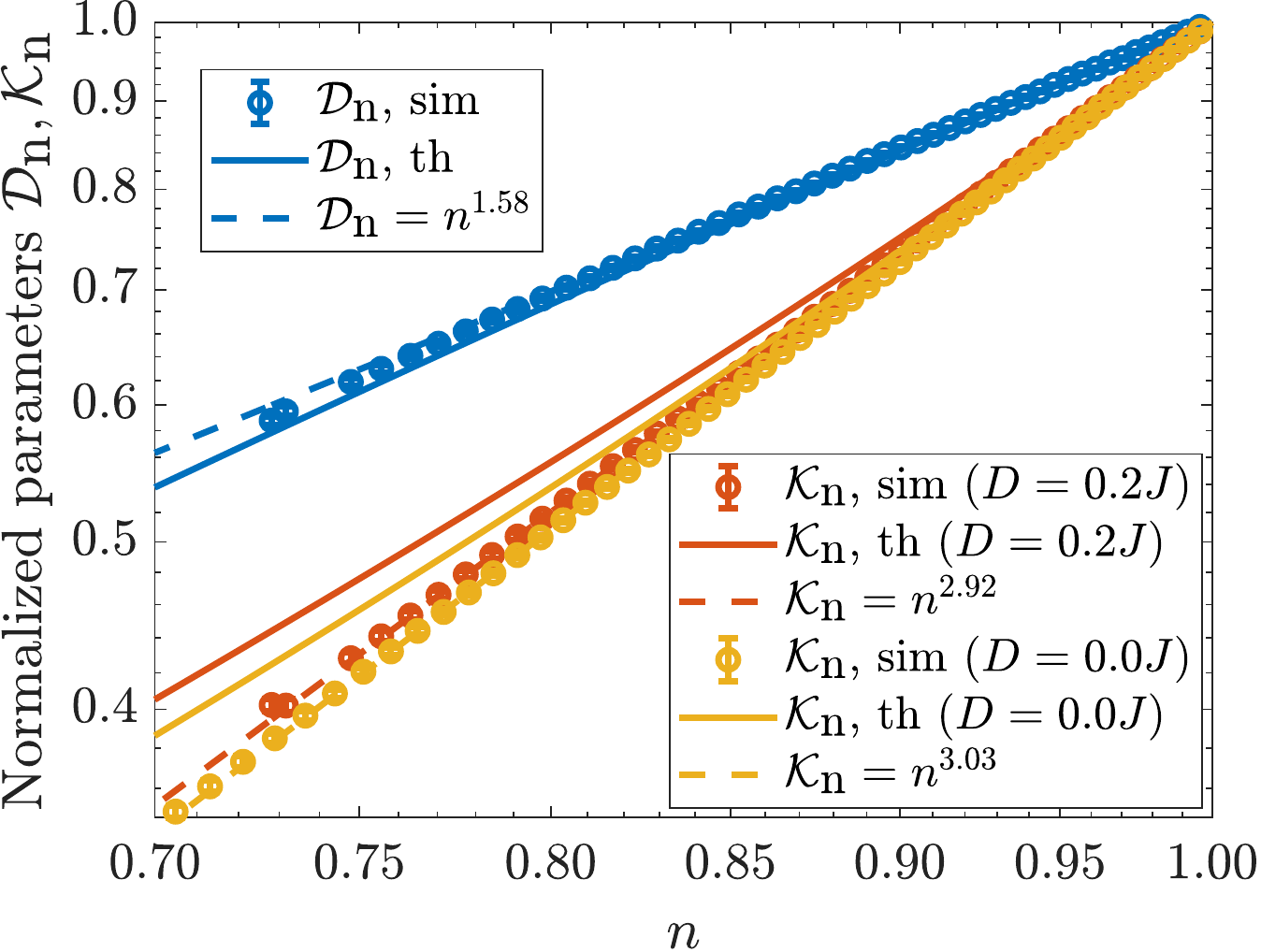}
    \caption{Dependence of the effective interaction parameters on the staggered magnetization $n$, equal to the sublattice magnetization $\left<S^{z}\right>$ in this system. All quantities are normalized to their zero-temperature value. Results of the numerical simulations (symbols) are compared to Green's function theory calculations (lines). Simulation data were obtained by fitting the functions in Eq.~\eqref{eqn32} and \eqref{eqn33} to the simulated frequencies; error bars denote the error of this fit. Dashed lines show a low-temperature power-law fit to the simulation data. The atomistic model parameters are $K=0.1J$ and $D=0.2J$ for the blue and orange curves, $D=0.0J$ for the yellow curves.\label{fig3}}
\end{figure}

The scaling of the parameters in the magnon spectrum with the staggered magnetization $n$ is shown in Fig.~\ref{fig3}, displaying a power-law behavior as discussed in Secs.~\ref{sec2b} and \ref{sec2c}. The explicit temperature dependence is illustrated in Fig.~\ref{fig3a} in Appendix~\ref{appendixA} for comparison. These results confirm the reliability of the Green's function method in predicting the simulation results. The Dzyaloshinsky--Moriya interaction $\mathcal{D}$ decreases slower in temperature than the anisotropy $\mathcal{K}$, as discussed in Sec.~\ref{sec2b} for the general case. The temperature dependence of the Heisenberg term $\mathcal{J}$ is identical to that of the Dzyaloshinsky--Moriya interaction in Green's function theory and agrees with it in the simulations within error bars; therefore, it is omitted from the figure. %The dependence on the order parameter, represented by the staggered magnetization $n$, may also be expressed as a power law. 
Based on a fit to the simulation data, the scaling exponent is $1.58$ for the Dzyaloshinsky--Moriya interaction, decreased by the correction $\varepsilon_{\textrm{inter}}=0.42$ compared to the uncorrelated value. 
The scaling exponent agrees with the value of $1.54-1.57$ obtained for the ferromagnetic case in Ref.~\cite{Rozsa}.
For the anisotropy, an exponent of $3.03$ is obtained without Dzyaloshinsky--Moriya interaction, rather close to the well-known power law predicting an exponent of $3$~\cite{Callen2}. In the presence of the Dzyaloshinsky--Moriya interaction, the exponent is slightly reduced to $2.92$, i.e., there is an additional positive contribution to the temperature dependence of the uniaxial anisotropy due to the Dzyaloshinsky--Moriya interaction. %Despite the slightly increased anisotropy, the presence of the Dzyaloshinsky--Moriya interaction does not stabilize antiferromagnetic order up to higher temperatures: Green's function theory calculations predict $k_{\textrm{B}}T_{\textrm{c}}\approx 0.84J$ for $D=0.2J$ and $k_{\textrm{B}}T_{\textrm{c}}\approx 0.88J$ for $D=0.0J$. Most likely this can be attributed to the reduced spin-wave gap created by the Dzyaloshinsky--Moriya interaction shown in Fig.~\ref{fig2}, which gap is essential for the stabilization of long-range order in two-dimensional systems.

\begin{figure}
    \centering
    \includegraphics[width = \columnwidth]{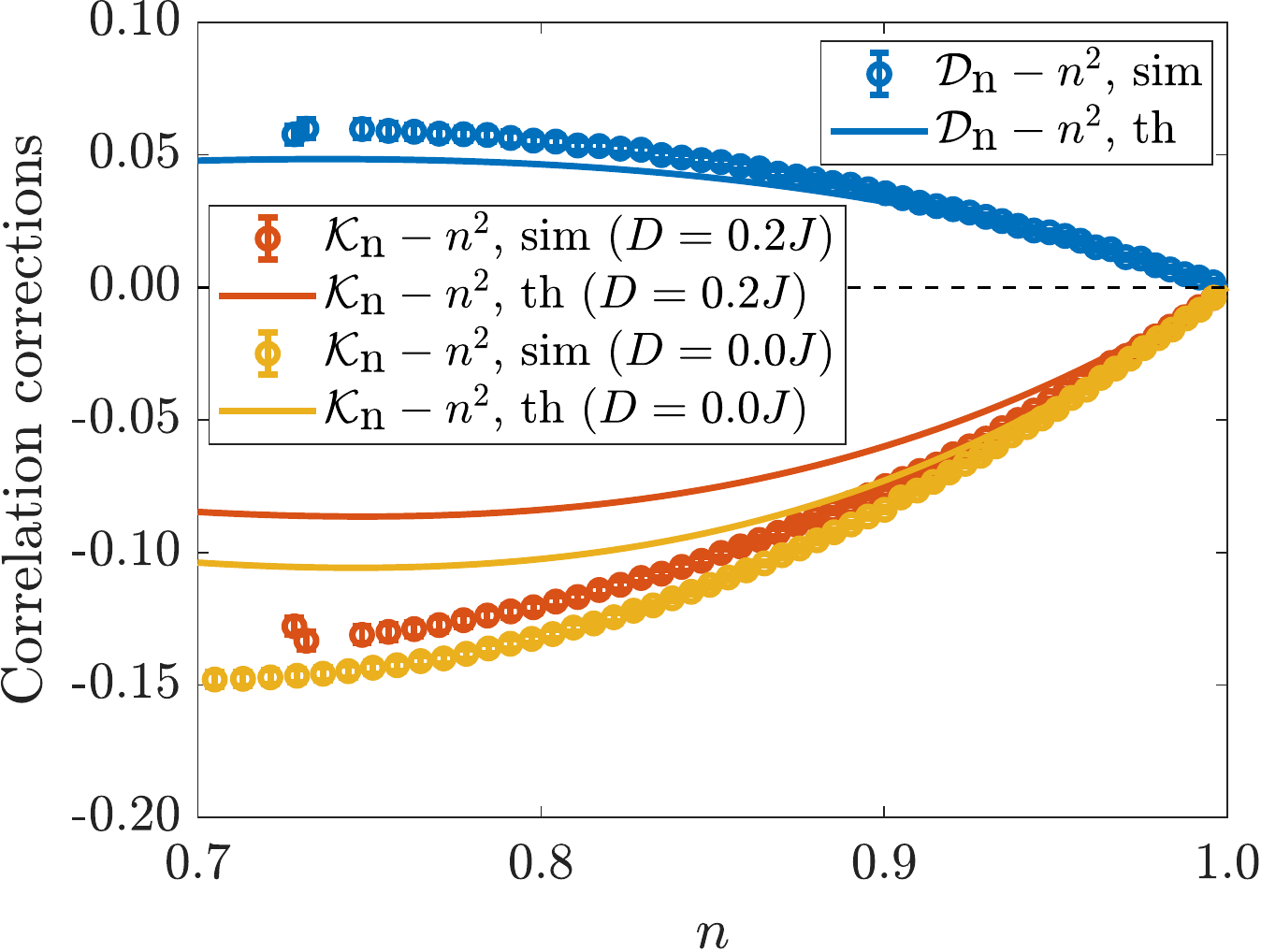}
    \caption{Correlation correction to the effective interaction parameters as a function of the staggered magnetization $n$. Data and notations are identical to Fig.~\ref{fig3}, apart from subtracting $n^{2}$ from the normalized parameters as indicated in the legend.\label{fig4}}
\end{figure}

The accuracy of the decoupling scheme may be better visualized after subtracting $n^{2}$ from the normalized parameters, leaving only the correlation corrections shown in Fig.~\ref{fig4}. Note that in the random-phase approximation obtained for $\alpha_{0}=0$, the curves would be zero as indicated by the dashed line in the figure. Comparing Figs.~\ref{fig3} and \ref{fig4}, it is clear that the correlation corrections are not negligible, contributing around $10\%$ of the total value of the Dzyaloshinsky--Moriya interaction and around $50\%$ of the total anisotropy at the highest simulated temperatures. As mentioned earlier, for $D=0$ the correction to the anisotropy will be $n^3-n^2$, i.e., it results in the Callen--Callen power law~\cite{Callen2}. The corrections are positive for the exchange and negative for the anisotropy terms as mentioned above, leading to increased and decreased effective exponents, respectively. While in this plot the deviations between Green's function theory and the simulations become apparent, even for the anisotropy terms the analytical description reproduces about $2/3$ of the corrections observed in the simulations. The accuracy appears to be higher for the Dzyaloshinsky--Moriya interaction itself and its correction to the anisotropy (difference of the orange and yellow lines).

\section{Conclusion}

We applied Green's function theory to calculate the magnon frequencies in two-sublattice antiferromagnetically aligned systems, and to determine the temperature dependence of the interaction parameters in the magnon spectrum. We found that transversal spin correlations stabilize the Heisenberg and Dzyaloshinsky--Moriya exchange interactions against thermal fluctuations, but induce a faster decay of the anisotropy terms with the temperature. The Dzyaloshinsky--Moriya interaction also contributes to the uniaxial anisotropy term via the spin correlations, increasing its value at finite temperature in contrast to the typical decrease. We obtained good agreement between the predictions of the theory and Monte Carlo simulations performed on a square lattice, where the correlations play a pronounced role due to the reduced dimensionality.

Remarkably, these observations do not simply qualitatively agree with previous calculations for ferromagnets~\cite{Bastardis,Rozsa,Evans}, but a mathematical correspondence can also be established. The self-consistency equations may be exactly transformed into each other in the classical limit when reversing the magnetization direction on one sublattice simultaneously with the sign of all intersublattice coupling terms. If the intrasublattice interactions are identical, the sublattice magnetizations and consequently the total and staggered magnetizations show precisely the same temperature dependence, even if the magnetic moments on the sublattices are different. Therefore, the scaling relations of the inter- and intrasublattice coupling terms discussed here can also be applied to ferromagnets with nearest-neighbor and next-nearest-neighbor interactions.
%These results agree with previous calculations for ferromagnets~\cite{Bastardis,Rozsa,Evans}. The agreement is not simply qualitative: the self-consistency equations may be exactly transformed into each other in the classical limit when reversing the magnetization direction on one sublattice simultaneously with the sign of all intersublattice coupling terms. If the intrasublattice interactions are identical, the sublattice magnetizations and consequently the total and staggered magnetizations show precisely the same temperature dependence, even if the magnetic moments on the sublattices are different.

The calculated temperature dependence of the parameters are fundamental for the development of multiscale models connecting first-principles spin-model parameters to finite-temperature mesoscopic computational approaches, such as micromagnetism or the Landau--Lifshitz--Bloch equation. Most of the multiscale approaches proposed so far
rely on an intermediate step based on classical spin-model simulations, which could be replaced by the considerably more efficient semi-analytical expression presented here. 
Multiscale methods would be able to access the dynamics of and the phase transitions in antiferromagnetically aligned systems, for example for a realistic and computationally efficient description of all-optical ultrafast switching processes in ferrimagnets~\cite{Raposo2022} 
or of magnetic domain wall motion in antiferromagnets~\cite{Hirst2022}.

Deviations in the equilibrium parameters from single-sublattice ferromagnets are expected to be observed in systems where the intrasublattice terms are not equivalent, such as ferrimagnets with a compensation point, or particularly when quantum effects are taken into account. Validating the predictions of Green's function theory in the quantum limit would require comparisons with classical spin-model simulations augmented by a semi-quantum thermostat~\cite{Barker2019} or with renormalized heat-bath temperatures~\cite{Evans2015}, or to quantum spin-model simulations based on quantum Monte Carlo~\cite{Sandvik2007} or tensor-product states~\cite{Cirac_2009}.  
The multi-scale quantum approach would be completed by using the calculated temperature-dependent parameters in the quantum version  
of the Landau-Lishitz-Bloch equation~\cite{Nieves2014}.

\section*{Acknowledgments}

L. R. gratefully acknowledges funding by the National Research, Development, and Innovation Office (NRDI) of Hungary under Project Nos. K131938 and FK142601, by the Ministry of Culture and Innovation and the National Research, Development and Innovation Office within the Quantum Information National Laboratory of Hungary (Grant No. 2022-2.1.1-NL-2022-00004), and by the Young Scholar Fund at the University of Konstanz.
U. A. gratefully acknowledges support by grant PID2021-122980OB-C55 and the grant RYC-2020-030605-I funded by MCIN/AEI/10.13039/501100011033 and by "ERDF A way of making Europe" and "ESF Investing in your future".

\appendix

\section{Derivation of the self-consistency equations\label{appendixA}}

The dynamics of the spin system is generated by the Poisson brackets
\begin{eqnarray}
\left\{S_{ir}^{\alpha},S_{js}^{\beta}\right\}&=&-\frac{\gamma}{\mu_{r}}\varepsilon^{\alpha\beta\gamma}\delta_{ij}\delta_{rs}S_{ir}^{\gamma}\label{eqnA1}
\end{eqnarray}
in the classical limit, where $i,j$ are lattice indices, $r,s$ are sublattice indices and $\alpha,\beta,\gamma$ are Cartesian indices as in the main text. In the quantum case, the Poisson brackets $\{S_{ir}^{\alpha},S_{js}^{\beta}\}$ have to be replaced by the commutators as $ -\textrm{i}\hbar^{-1}[S_{ir}^{\alpha},S_{js}^{\beta}]$, and $\gamma/\mu_{r}$ has to be replaced by $\hbar^{-1}$ in Eq.~\eqref{eqnA1} and the following expressions. Note that in the quantum limit we introduced a sign change compared to the conventional commutation relations, since the $S_{ir}^{\alpha}$ operators represent the dimensionless magnetic moments of electrons which are antiparallel to the angular momenta.

The time-dependent Green's function is defined as
\begin{eqnarray}
G_{i;j}^{r\alpha;s\beta}\left(t;u\right)=\theta\left(t\right)\left<\left\{S_{ir}^{\alpha}\left(t\right),\textrm{e}^{uS_{js}^{z}\left(0\right)}S_{js}^{\beta}\left(0\right)\right\}\right>,\label{eqnA2}
\end{eqnarray}
where $\theta\left(t\right)$ is the Heaviside function, $\left<\right>$ denotes averaging in thermal equilibrium, and $u$ is a real parameter. Here, $\alpha$ and $\beta$ will primarily denote the ladder operator indices $+$ and $-$, but unless they are explicitly specified the expressions are also valid for the Cartesian indices. The Green's function satisfies the equation of motion
\begin{eqnarray}
\partial_{t}G_{i;j}^{r\alpha;s\beta}=&&\delta\left(t\right)\left<\left\{S_{ir}^{\alpha},\textrm{e}^{uS_{js}^{z}}S_{js}^{\beta}\right\}\right>\nonumber
\\
&&+\theta\left(t\right)\!\left<\!\left\{\!\left\{S_{ir}^{\alpha}\left(t\right),H\right\},\textrm{e}^{uS_{js}^{z}\left(0\right)}S_{js}^{\beta}\left(0\right)\!\right\}\!\right>\!\!.\label{eqnA3}
\end{eqnarray}

The second term on the right-hand side of Eq.~\eqref{eqnA3} introduces higher-order Green's functions, which will be decoupled using
\begin{align}
G_{ij;k}^{r- sz;t\alpha}\approx\left<S_{js}^{z}\right>G_{i;k}^{r-;t\alpha}-\alpha^{s+r-}\left<S_{js}^{+}S_{ir}^{-}\right>G_{j;k}^{s-;t\alpha}.\label{eqnA4}
\end{align}
Here we took advantage of the rotational symmetry of the system: only such expectation values are considered in the decoupling which are rotationally invariant around the $z$ axis, namely $\left<S_{ir}^{z}\right>\equiv\left<S_{r}^{z}\right>$ which is the same at all sites in the sublattice due to the translational invariance of the ground state, and $\left<S_{js}^{+}S_{ir}^{-}\right>$ that is replaced by half of the anticommutator in the quantum limit. The decoupling coefficients are chosen as $\alpha^{s+r-}=\left<S_{r}^{z}\right>\alpha_{0}^{sr}$, with $\alpha_{0}^{sr}=1/2$ in the classical and $\alpha_{0}^{sr}=1/\left(2S_{r}S_{s}\right)$ in the quantum case. This choice of the decoupling parameters will be motivated later. 

We introduce the transformed coordinate system with the sublattice spins pointing along the local $z$ direction as discussed in the main text, and perform temporal and spatial Fourier transformation via $\partial_{t}\rightarrow -\textrm{i}\omega$ and
\begin{eqnarray}
\tilde{G}_{\boldsymbol{q}}^{r\alpha;s\beta}&=&\frac{1}{N_{\textrm{c}}}\sum_{\boldsymbol{R}_{i}-\boldsymbol{R}_{j}}\textrm{e}^{-\textrm{i}\boldsymbol{q}\left(\boldsymbol{R}_{i}-\boldsymbol{R}_{j}\right)}\tilde{G}_{ij}^{r\alpha;s\beta},\label{eqnA5}
\\
%\left<\tilde{S}_{-\boldsymbol{q}r}^{\alpha}\tilde{S}_{\boldsymbol{q}s}^{\beta}\right>&=&\frac{1}{N_{\textrm{c}}}\sum_{\boldsymbol{R}_{i}-\boldsymbol{R}_{j}}\textrm{e}^{-\textrm{i}\boldsymbol{q}\left(\boldsymbol{R}_{i}-\boldsymbol{R}_{j}\right)}\left<\tilde{S}_{ir}^{\alpha}\tilde{S}_{js}^{\beta}\right>.\label{eqnA6}
\left<\tilde{S}_{-\boldsymbol{q}r}^{\alpha}\tilde{S}_{\boldsymbol{q}s}^{\beta}\right>&=&\frac{1}{N_{\textrm{c}}}\sum_{\boldsymbol{R}_{i}-\boldsymbol{R}_{j}}\textrm{e}^{-\textrm{i}\boldsymbol{q}\left(\boldsymbol{R}_{i}-\boldsymbol{R}_{j}\right)}\left<\tilde{S}_{jr}^{\alpha}\tilde{S}_{is}^{\beta}\right>.\label{eqnA6}
\end{eqnarray}

Following the decoupling and the Fourier transformation, Eq.~\eqref{eqnA3} reads
\begin{eqnarray}
\omega\tilde{G}_{\boldsymbol{q}}^{r(2);t\alpha}=\frac{1}{2\pi N_{\textrm{c}}}\frac{\gamma}{\mu_{r}}\delta_{rt}\tilde{\Theta}^{r(2);t\alpha}+\sum_{s}\frac{\gamma}{\mu_{r}}\tilde{\Gamma}^{rs}_{\boldsymbol{q}}\tilde{G}_{\boldsymbol{q}}^{s(2);t\alpha},\label{eqnA7}
\end{eqnarray}
which is related to the linearized equation of motion in Eq.~\eqref{eqn1e}. The additional inhomogeneous term contains
\begin{eqnarray}
\tilde{\Theta}^{r\beta;t\alpha}=\textrm{i}\frac{\mu_{r}}{\gamma}\left<\left\{\tilde{S}_{ir}^{\beta},\textrm{e}^{u\tilde{S}_{ir}^{z}}\tilde{S}_{ir}^{\alpha}\right\}\right>,\label{eqnA8}
\end{eqnarray}
and the $\tilde{\Gamma}^{rs}_{\boldsymbol{q}}$ coefficients are introduced in Eq.~\eqref{eqn1g} in the main text, with the components given by
\begin{align}
\tilde{\Gamma}^{AA}_{\boldsymbol{q}}=&\mathfrak{J}_{\boldsymbol{0}}^{AA}+\mu_{A}B^{z}-\mathfrak{J}_{\boldsymbol{0}}^{AB}-\mathfrak{J}_{\boldsymbol{q}}^{'AA}-2\alpha_{0}\nonumber
\\
&\times\sum_{\boldsymbol{q}'}\left[\left(\mathfrak{J}_{\boldsymbol{q}-\boldsymbol{q}'}^{AA}-\mathfrak{J}_{\boldsymbol{q}'}^{'AA}\right)\Phi^{AA}_{\boldsymbol{q}'}\left<\tilde{S}_{A}^{z}\right>+\mathfrak{J}_{\boldsymbol{q}'}^{'AB}\Phi^{AB}_{\boldsymbol{q}'}\left<\tilde{S}_{B}^{z}\right>\right],\label{eqn4}
\\
\tilde{\Gamma}^{AB}_{\boldsymbol{q}}=&\mathfrak{J}_{\boldsymbol{q}}^{'AB}+2\alpha_{0}\sum_{\boldsymbol{q}'}\mathfrak{J}_{\boldsymbol{q}-\boldsymbol{q}'}^{AB}\Phi^{BA}_{\boldsymbol{q}'}\left<\tilde{S}_{A}^{z}\right>,\label{eqn5}
\\
\tilde{\Gamma}^{BA}_{\boldsymbol{q}}=&-\mathfrak{J}_{\boldsymbol{q}}^{'BA}-2\alpha_{0}\sum_{\boldsymbol{q}'}\mathfrak{J}_{\boldsymbol{q}-\boldsymbol{q}'}^{BA}\Phi^{AB}_{\boldsymbol{q}'}\left<\tilde{S}_{B}^{z}\right>,\label{eqn6}
\\
\tilde{\Gamma}^{BB}_{\boldsymbol{q}}=&-\mathfrak{J}_{\boldsymbol{0}}^{BB}+\mu_{B}B^{z}+\mathfrak{J}_{\boldsymbol{0}}^{BA}+\mathfrak{J}_{\boldsymbol{q}}^{'BB}+2\alpha_{0}\nonumber
\\
&\times\sum_{\boldsymbol{q}'}\left[\left(\mathfrak{J}_{\boldsymbol{q}-\boldsymbol{q}'}^{BB}-\mathfrak{J}_{\boldsymbol{q}'}^{'BB}\right)\Phi^{BB}_{\boldsymbol{q}'}\left<\tilde{S}_{B}^{z}\right>+\mathfrak{J}_{\boldsymbol{q}'}^{'BA}\Phi^{BA}_{\boldsymbol{q}'}\left<\tilde{S}_{A}^{z}\right>\right],\label{eqn7}
\end{align}

We again used the short-hand notation $\tilde{S}_{\boldsymbol{q}r}^{(2)}\in\left\{\tilde{S}_{\boldsymbol{q}A}^{-},\tilde{S}_{\boldsymbol{q}B}^{+}\right\}$. As in linear spin-wave theory in Eq.~\eqref{eqn1e}, the corresponding equations for the $\tilde{S}_{\boldsymbol{q}r}^{(1)}\in\left\{\tilde{S}_{\boldsymbol{q}A}^{+},\tilde{S}_{\boldsymbol{q}B}^{-}\right\}$ spin components decouple from Eq.~\eqref{eqnA7}. The equations for $\tilde{S}_{\boldsymbol{q}r}^{(1)}$ yield the other branch of the spin-wave dispersion relation shown in Fig.~\ref{fig2}, but they need not be solved separately since they are connected to the $\omega_{\boldsymbol{q}}^{\pm}$ frequencies by particle-hole symmetry. Equation~\eqref{eqnA7} is solved as
\begin{eqnarray}
\tilde{G}_{\boldsymbol{q}}^{r(2);t\alpha}=\left(\omega-\frac{\gamma}{\mu_{r}}\tilde{\Gamma}^{rs}_{\boldsymbol{q}}\right)^{-1}\frac{1}{2\pi N_{\textrm{c}}}\frac{\gamma}{\mu_{s}}\delta_{st}\tilde{\Theta}^{s(2);t\alpha},\label{eqnA9}
\end{eqnarray}
where the inverse matrix has poles at the real frequencies $\omega_{\boldsymbol{q}}^{\pm}$ given in Eq.~\eqref{eqn13}. These poles may be used to evaluate the correlation functions via the spectral theorem (cf. Eq.~\eqref{eqn1i}),
\begin{align}
&\left<\textrm{e}^{u\tilde{S}_{js}^{z}\left(0\right)}\tilde{S}_{js}^{\beta}\left(0\right)\tilde{S}_{ir}^{\alpha}\left(t\right)\right>\nonumber
\\
&=\lim_{\varepsilon\rightarrow 0}\int
-2\textrm{Im}\left(\tilde{G}_{i;j}^{r\alpha;s\beta}\left(\omega+\textrm{i}\varepsilon;u\right)\right)n\left(\omega\right)\textrm{e}^{-\textrm{i}\omega t}\textrm{d}\omega,\label{eqnA10}
\end{align}
where $n\left(\omega\right)$ is the function introduced in the main text, corresponding to the occupation number in units of action for $\omega>0$. Equation~\eqref{eqnA10} results in
\begin{align}
\Phi^{AA}_{\boldsymbol{q}}=&\frac{\gamma}{\mu_{A}}\frac{1}{2N_{\textrm{c}}}\left(\frac{1}{\nu_{\boldsymbol{q}}}\left[n\left(\omega^{+}_{\boldsymbol{q}}\right)-n\left(\omega^{-}_{\boldsymbol{q}}\right)\right]+n\left(\omega^{+}_{\boldsymbol{q}}\right)-n\left(-\omega^{-}_{\boldsymbol{q}}\right)\right),\label{eqn8}
\\
\Phi^{AB}_{\boldsymbol{q}}=&\frac{\gamma}{\mu_{A}}\frac{\gamma}{\mu_{B}}\frac{1}{N_{\textrm{c}}}\frac{1}{\omega^{+}_{\boldsymbol{q}}-\omega^{-}_{\boldsymbol{q}}}\tilde{\Gamma}^{BA}_{\boldsymbol{q}}\left(n\left(\omega^{+}_{\boldsymbol{q}}\right)-n\left(\omega^{-}_{\boldsymbol{q}}\right)\right),\label{eqn9}
\\
\Phi^{BA}_{\boldsymbol{q}}=&-\frac{\gamma}{\mu_{A}}\frac{\gamma}{\mu_{B}}\frac{1}{N_{\textrm{c}}}\frac{1}{\omega^{+}_{\boldsymbol{q}}-\omega^{-}_{\boldsymbol{q}}}\tilde{\Gamma}^{AB}_{\boldsymbol{q}}\left(n\left(\omega^{+}_{\boldsymbol{q}}\right)-n\left(\omega^{-}_{\boldsymbol{q}}\right)\right),\label{eqn10}
\\
\Phi^{BB}_{\boldsymbol{q}}=&\frac{\gamma}{\mu_{B}}\frac{1}{2N_{\textrm{c}}}\left(\frac{1}{\nu_{\boldsymbol{q}}}\left[n\left(\omega^{+}_{\boldsymbol{q}}\right)-n\left(\omega^{-}_{\boldsymbol{q}}\right)\right]-n\left(\omega^{+}_{\boldsymbol{q}}\right)+n\left(-\omega^{-}_{\boldsymbol{q}}\right)\right),\label{eqn11}
\end{align}
for $u=0$. Here Eq.~\eqref{eqn1h} was used to introduce the $\Phi^{rs}_{\boldsymbol{q}}$ quantities, and $\tilde{\Theta}^{A-;A+}\left(u=0\right)=2\left<S_{A}^{z}\right>$ and $\tilde{\Theta}^{B+;B-}\left(u=0\right)=-2\left<S_{B}^{z}\right>$ were substituted based on the Poisson brackets. In the classical limit, these equations simplify to 
 %Substituting the expression for $n\left(\omega^{\pm}_{\boldsymbol{q}}\right)$ simplifies Eqs.~\eqref{eqn8}-\eqref{eqn11} to
\begin{eqnarray}
\Phi^{AA}_{\boldsymbol{q}}&=&\frac{1}{N_{\textrm{c}}}\frac{k_{\textrm{B}}T}{\textrm{det}\:\tilde{\Gamma}_{\boldsymbol{q}}}\tilde{\Gamma}^{BB}_{\boldsymbol{q}},\label{eqn14}
\\
\Phi^{AB}_{\boldsymbol{q}}&=&-\frac{1}{N_{\textrm{c}}}\frac{k_{\textrm{B}}T}{\textrm{det}\:\tilde{\Gamma}_{\boldsymbol{q}}}\tilde{\Gamma}^{BA}_{\boldsymbol{q}},\label{eqn15}
\\
\Phi^{BA}_{\boldsymbol{q}}&=&\frac{1}{N_{\textrm{c}}}\frac{k_{\textrm{B}}T}{\textrm{det}\:\tilde{\Gamma}_{\boldsymbol{q}}}\tilde{\Gamma}^{AB}_{\boldsymbol{q}},\label{eqn16}
\\
\Phi^{BB}_{\boldsymbol{q}}&=&-\frac{1}{N_{\textrm{c}}}\frac{k_{\textrm{B}}T}{\textrm{det}\:\tilde{\Gamma}_{\boldsymbol{q}}}\tilde{\Gamma}^{AA}_{\boldsymbol{q}},\label{eqn17}
\end{eqnarray}
with $\textrm{det}\:\tilde{\Gamma}_{\boldsymbol{q}}=\tilde{\Gamma}^{AA}_{\boldsymbol{q}}\tilde{\Gamma}^{BB}_{\boldsymbol{q}}-\tilde{\Gamma}^{AB}_{\boldsymbol{q}}\tilde{\Gamma}^{BA}_{\boldsymbol{q}}$. These equations are summarized in Eq.~\eqref{eqn1j}. Note that $\Phi_{\boldsymbol{q}}^{rs}$ is symmetrized by the anticommutator in the quantum case in the convention used here, and the components are given by Eqs.~\eqref{eqn8}-\eqref{eqn11}. The sublattice-diagonal part of Eq.~\eqref{eqnA10} also yields a differential equation in $u$, which is of the same form as the one described in Ref.~\cite{Callen} in the quantum case and in Ref.~\cite{Rozsa} in the classical limit; the solution of this equation with the appropriate boundary conditions gives the final equation \eqref{eqn18} or \eqref{eqn19} required for self-consistency.

The intersublattice terms have to satisfy $\Phi_{\boldsymbol{q}}^{AB}=\left<\tilde{S}_{-\boldsymbol{q}A}^{+}\tilde{S}_{\boldsymbol{q}B}^{+}\right>=\left<\tilde{S}_{-\left(-\boldsymbol{q}\right)B}^{-}\tilde{S}_{-\boldsymbol{q}A}^{-}\right>^{*}=\Phi_{-\boldsymbol{q}}^{BA*}$. When reintroducing the general decoupling coefficients $\alpha^{s+r-}$ from Eq.~\eqref{eqnA4} in Eqs.~\eqref{eqn9} and \eqref{eqn10}, this leads to the constraint $\left<\tilde{S}_{B}^{z}\right>\alpha^{B-A-}=\left<\tilde{S}_{A}^{z}\right>\alpha^{A+B+}$, as discussed in Ref.~\cite{Anderson}. This constraint is satisfied by the choice used in the main text and described after Eq.~\eqref{eqnA4}. However, it also allows for using different decoupling schemes for the intrasublattice and the intersublattice terms.

\begin{figure}
    \centering
    \includegraphics[width = \columnwidth]{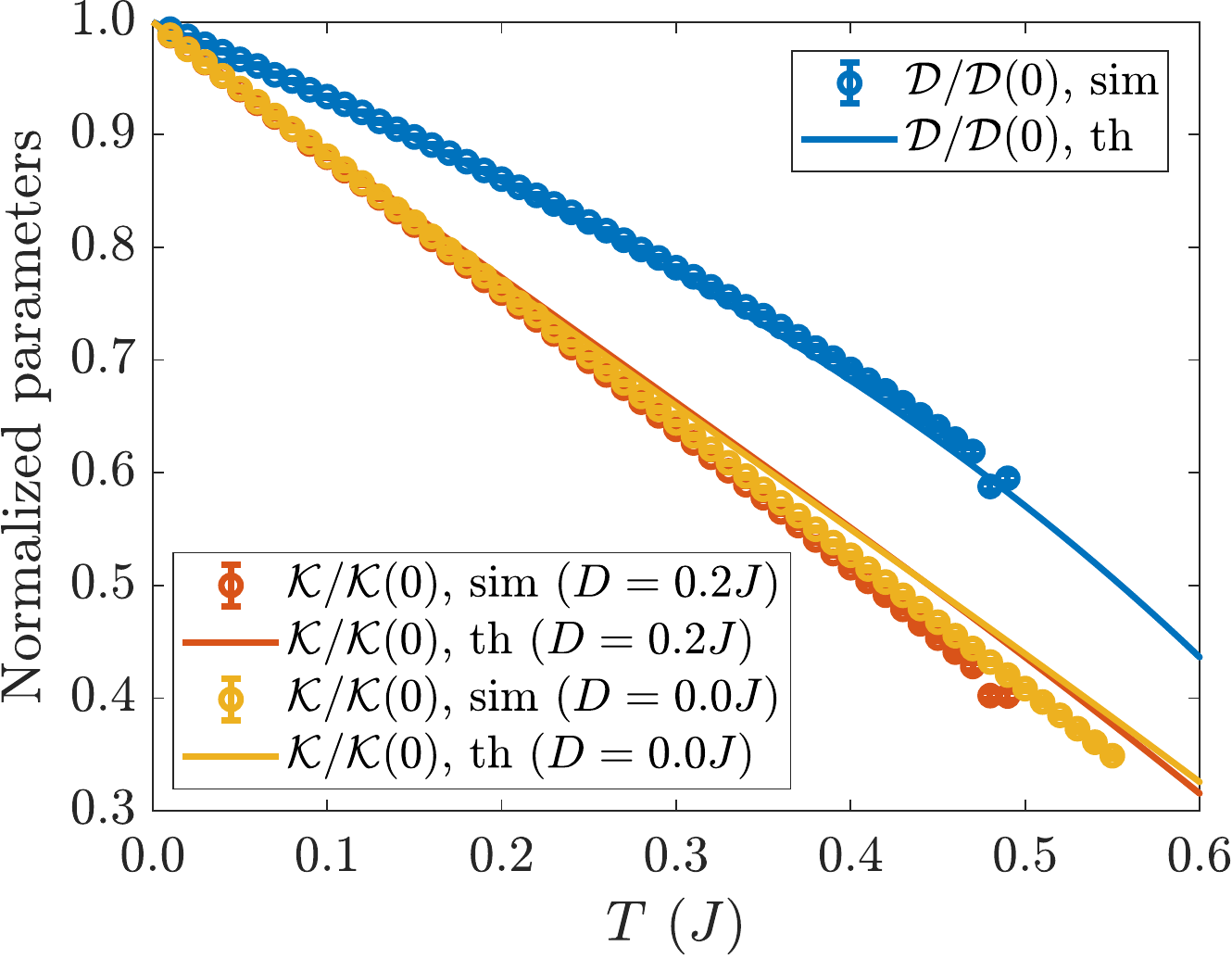}
    \caption{Dependence of the effective interaction parameters on the temperature. All quantities are normalized to their zero-temperature value. Results of the numerical simulations (symbols) are compared to Green's function theory calculations (lines). Simulation data were obtained by fitting the functions in Eqs.~\eqref{eqn32} and \eqref{eqn33} to the simulated frequencies; error bars denote the error of this fit. The atomistic model parameters are $K=0.1J$ and $D=0.2J$ for the blue and orange curves, $D=0.0J$ for the yellow curves.\label{fig3a}}
\end{figure}

\end{document}